\documentclass[12pt,a4paper,english,nofootinbib,,superscriptaddress]{revtex4}
\usepackage{lmodern}

\usepackage[T1]{fontenc}
\usepackage[latin9]{inputenc}
\usepackage{babel}
\usepackage{color}
\usepackage{amsmath}
\usepackage{graphicx}
\usepackage{amssymb}
\usepackage[unicode=true, pdfusetitle,
bookmarks=true,bookmarksnumbered=false,bookmarksopen=false,
breaklinks=false,pdfborder={0 0 1},backref=false,colorlinks=false]
{hyperref}
\usepackage{amsmath}
\usepackage{amssymb}
\usepackage{float}

\def\b{\begin{equation}} \def\e{\end{equation}}
\def\bd{\begin{displaystyle}} \def\ed{\end{displaystyle}}
\def\ba{\begin{array}} \def\ea{\end{array}}

\def\bee{\begin{enumerate}}
	\def\eee{\end{enumerate}}

\def\1{\mbox{I\hspace{-.15em}1}}

\def\b{\begin{equation}}
\def\e{\end{equation}}
\def\bee{\begin{enumerate}}
	\def\eee{\end{enumerate}}

\makeatletter
\usepackage{latexsym}\usepackage{bm}

\makeatother
\begin{document}
	
	$$$$
	
	\title{\Large\textbf{Quantum corrected thermodynamics of nonlinearly charged BTZ black holes in massive gravity's rainbow}}

	\author{M. Dehghani}\email{m.dehghani@razi.ac.ir}
		
	\affiliation{Department of Physics, Razi University, Kermanshah, Iran}
	\author{B. Pourhassan}\email{b.pourhassan@umz.ac.ir}
		
	\affiliation{School of Physics, Damghan University, Damghan,
3671641167, Iran}
	\begin{abstract}
		
		$$$$
		
		\begin{center}
		\textbf{ABSTRACT}
     	\end{center}

     In this paper, we consider three-dimensional massive gravity's rainbow and obtain black hole solutions in three different cases of Born-Infeld, logarithmic, and exponential theories of nonlinear electrodynamics. We discuss the horizon structure and geometrical properties. Then, we study thermodynamics of these models by considering the first-order quantum correction effects, which appears as a logarithmic term in the black hole entropy. We discuss such effects on the black hole stability and phase transitions. We find that, due to the quantum corrections, the second-order phase transition happens in Born-Infeld and logarithmic models. We obtain the modified first law of black hole thermodynamics in the presence of logarithmic corrections.\\
	
\end{abstract}
	
	\maketitle

\newpage

\setcounter{equation}{0}
\section{\textbf{Introduction}}
Several observations indicate that our Universe expands with positive acceleration \cite{001, 002, 003, 004, 005, 006}. This accelerated expansion can be explained by considering dark energy which includes several models like cosmological constant, scalar fields and related fields \cite{007, 008, 009, 0010, 0011}, or extended Chaplygin gas \cite{ECG}. Also, it can be explained by use of the so-called modified theories of gravity such as $f(R)$ gravity \cite{0012}, $f(R, T)$ gravity \cite{0013}, $f(R, L)$ gravity \cite{0014, 0015}, Gauss-Bonnet and Lovelock gravities \cite{0016, 0017}, Weyl gravity \cite{0018}, and Ho\v{r}ava-Lifshitz gravity \cite{HL}.\\
Detection of the gravitational waves, as one of the interesting predictions of Einstein's original theory of gravity, by LIGO and Virgo experiments confirmed the validity of this theory \cite{1, 2, 3}. This theory, which is known as the general relativity, also predicts existence of the massless gravitons as the mediators of gravitational interactions. Massive gravity theory, as an alternative theory of gravity, is a natural extension of general relativity to include massive gravitons. It was initially proposed by Fierz-Pauli base on the fact that the gravitons may have a nonzero amount of mass \cite{fp, hi}. The ghost free and physically reasonable nonlinear Fierz-Pauli massive theory was finalized by de Rham, Gabadadze and Tolley (dRGT) in 2010 \cite{dRGT1, dRGT2}. Interestingly, the dRGT massive gravity theory can describe the accelerated expansion of the Universe without need of dark matter and dark energy components \cite{12, 11, 10}. Also, it was already used to study black holes as well as cosmology \cite{BH1, BH2, chabab, Zhang, hendihep, CO1}.

It is well-known that gravity's rainbow/or doubly general relativity, as the ultraviolet extension of general relativity, has been proposed to address the failures of Einstein's gravity theory in the high energy regime. In high energy physics, where the energies are comparable with Planck energy, the usual energy-momentum relation must be promoted to the so-called modified dispersion relation. In terms of the rainbow functions $f(\varepsilon)$ and $g(\varepsilon)$ it is written in the following form  \cite{am1, 3drain, 4dRG}
\b E^2 f^2(\varepsilon)-p^2 g^2(\varepsilon)=m^2.\e
The energy scale $\varepsilon$  is defined as $\varepsilon=E/E_P$, in which $E_P$ is the Planck energy and $E$ is the energy of a test particle which probes the gravity. The temporal and spatial functions $f(\varepsilon)$ and $g(\varepsilon)$ are required to be equal to one in the limit $\varepsilon\rightarrow 0$. It is evident that in this limiting case, the modified dispersion relation reduces to the usual dispersion relation, and doubly general relativity recovers its infrared version. Doubly special relativity, as the high energy deformation of usual special relativity, by treating Planck energy and speed of light as two observer-independent quantities, preserves Lorentz invariance of modified dispersion relation. In this theory, the inertial observers are related via a set of modified Lorentz transformations \cite{am5, 4dn, 3dn}. Doubly general relativity is constructed out by generalization of doubly special relativity which, just like Einstein's gravity theory, includes gravity \cite{mag1, mag2}. An interesting property of this theory is that the geometry of space-time depends on the test particle's  energy. Thus, particles with different amounts of energy experience different geometries. Therefore, instead of a unique metric, there is a rainbow of metrics, and this theory is named rainbow gravity \cite{4dpower, 3dpower}. Gravity's rainbow has been the subject of many interesting research works, and some novel and important results have been produced by consideration of the energy-dependent space-times. Thermal stability of the black hole and non-singularity of the universe are the outstanding results of applying gravity's rainbow \cite{fa1, non1, fa2, non2, noz, non3}.

On the other hand, the nonlinear theory of electrodynamics was initially proposed for explaining the failures of Maxwell's electromagnetic theory. Appearance of the infinite field and self-energy at the position of point-like charged particles, and violation of conformal symmetry in the space-times with dimension non-equal to four are the most important challenges of Maxwell's electrodynamics \cite{mh, 3dCI, 4dCI}. The Born-Infeld electrodynamics is one of the most famous theories of nonlinear electrodynamics \cite{bin, bi1}. There are also other models of nonlinear electrodynamics like power-law, logarithmic, exponential, and quadratically-extended models \cite{hamidi1, log, exp, qe}. Using nonlinear electrodynamics in modified theories of gravity is interesting to study various black hole solutions \cite{ kr1, hamidi2, zeb, kazemi, epjp, hajkh}. In that case, the exact black hole solutions of the Einstein-massive gravity theory were obtained recently by using Maxwell and power-Maxwell theories. Then, by use of this method, the thermodynamics properties of the massive-BTZ black holes can be investigated \cite{3dmpl}.

In this paper, we would like to extend Ref. \cite{3dmpl} to the case of asymptotically massive-BTZ black holes by considering the three different cases of Born-Infeld (BI), logarithmic (LO), and exponential (EX) theories of nonlinear electrodynamics. This study can yield to better understanding of the massive gravity theory, and the impacts of rainbow functions on the thermodynamic stability of black holes. A black hole with negative specific heat is in the unstable phase. In such a situation, a black hole reduces its size due to the Hawking radiation. In that case, there are two possible scenarios for the final stage. In the first one, a black hole evaporates when all of its mass radiates. On the other hand, in the second scenario, a black hole can stop radiation and become stable. Perturbation theory suggests a logarithmic correction (at leading order) on the black hole entropy, as the black hole reduces its size due to the Hawking radiation. When the size is reduced more, the second order correction becomes also important. These are due to the thermal fluctuations which are significant for the tiny black holes \cite{NPB}. However, reducing the size more and more, the non-perturbative correction also becomes important which is an exponential correction term, and can be interpreted as a quantum effect \cite{cqg}. In this paper we deal with the leading order corrected entropy.\\

It is well-known that the black hole entropy is proportional to the horizon area,
\begin{equation}\label{s0}
S_{0}=\frac{\mathcal{A}}{4l_{p}^{2}},
\end{equation}
where, $\mathcal{A}$ is the horizon area and $l_{p}$  is the Planck length. Thermal fluctuations correct the black hole entropy and there are several ways to obtain such corrections. Recently, it has been suggested that correction terms are universal and have the following approximate shape \cite{2007.15401},
\begin{equation}\label{s}
S=S_{0}+\alpha \ln{S_{0}}+\frac{\gamma}{S_{0}}+\eta e^{-S_{0}}+...
\end{equation}
where, $\alpha$, $\gamma$ and $\eta$ are constant coefficients, and dots denote higher-order corrections. In this paper, we are interested in the case of logarithmic correction \cite{L1, L2, L3} to find its effect on the black hole thermodynamics, while exponential correction already discussed for other black hole solution \cite{exp}.\\
This paper is organized as follows. In the next section, we review the field equations and solve them in a spherically symmetric energy-dependent geometry. In section III, we calculate the impacts of rainbow functions on the black hole thermodynamic quantities and investigate the validity of the first law of the black hole thermodynamics. In section IV, we study local stability of the black holes by use of the canonical ensemble method. In section V, we consider logarithmic corrected entropy and study corrected thermodynamics of the models. Finally, in section VI, we give the conclusions.

\setcounter{equation}{0}
\section{\textbf{The field equations and solutions}}

The action of three-dimensional massive gravity given by \cite{3dmpl, hendiplb},
\b I=\frac{1}{16\pi}\int\sqrt{-g}\left[{\cal{R}}-2\Lambda+m_G^2\sum_{i=1}^4c_i{\cal{U}}_i+L({\cal{F}})\right]d^3x,\label{1} \e
where, ${\cal{R}}$ is the Ricci scalar, $L({\cal{F}})$, is the Lagrangian of nonlinear electrodynamics which is assumed as a function of Maxwell's invariant ${\cal{F}}=F^{ab}F_{ab}$, with $F_{ab}=\partial_a A_b-\partial_b A_a$ in terms of electromagnetic four-vector $A_a$. Also,  $\Lambda=-\ell^{-2}$ is the cosmological constant, and the parameter $m_G$ is related to the graviton mass. The $c_i$s are some constant coefficients, and ${\cal{U}}_i$s are polynomials of eigenvalues of the  $3\times3$ matrix ${\cal{K}}^a_{\;b}=\sqrt{g^{a c}f_{c b}}$, with $f_{ab}$ as a symmetric rank-two tensor. These polynomials are given by the following explicit relations \cite{cai, hendiprd},

\begin{eqnarray}\label{2}
{\cal{U}}_1&=& [{\cal{K}}],\nonumber\\
{\cal{U}}_2&=&[{\cal{K}}]^2-[{\cal{K}}^2],\nonumber\\
{\cal{U}}_3&=&[{\cal{K}}]^3-3[{\cal{K}}][{\cal{K}}^2]+2[{\cal{K}}^3],\nonumber\\
{\cal{U}}_4&=&[{\cal{K}}]^4-6[{\cal{K}}]^2[{\cal{K}}^2]+8[{\cal{K}}^3][{\cal{K}}]+3[{\cal{K}}^2]^2-6[{\cal{K}}^4].
\end{eqnarray}
Note that the notation $[{\cal{K}}]\equiv{\cal{K}}^a_a$ has been used, here.

We are interested in the following Lagrangian of nonlinear electrodynamics \cite{2012},
\b L({\cal{F}})=\left\{\begin{array}{ll}\label{3}
	4b^2\left(1-\sqrt{1+\frac{{\cal{F}}}{2b^2}}\right),\;\;(BI)\\\\
	
	-8b^2 \ln \left(1+\frac{{\cal{F}}}{8b^2}\right),\;\;(LO)\\\\
	
	-b^2\left(1-e^{-\frac{{\cal{F}}}{b^2}}\right),\;\;(EX)
\end{array} \right.\e
where, the constant $b$ is the nonlinearity parameter. The Lagrangian (\ref{3}) is reduced to the Lagrangian of Maxwell's classical electrodynamics at large $b$ ($b\rightarrow\infty$) limit (which is corresponding to small ${\cal{F}}$ ).

By varying the action (\ref{1}), with respect to different fields, one can obtain the following tensor and vector equations \cite{cai, hendiprd}
\b{\cal{R}}_{ab}-\frac{1}{2}{\cal{R}} g_{ab}+\Lambda g_{ab}+m^2_G
\chi_{ab}= \frac{1}{2}L({\cal{F}})
g_{ab}-2L'({\cal{F}})F_{ac}F^{\;\;c}_{b}, \label{4}\e
\b\partial_a\left[\sqrt{-g}L'({\cal{F}})F^{ab}\right]=0,\label{5}\e
for the gravitational and electromagnetic fields, respectively. Here prim means derivative with respect to the argument. The term $\chi_{ab}$, in equation (\ref{4}), is related to the massive gravity and is given by the following relation\cite{cai},

\begin{eqnarray}\label{6}
\chi_{ab}=&-&\frac{c_1}{2}\left({\cal{U}}_1g_{ab}-{\cal{K}}_{ab}\right)\nonumber\\
&-&\frac{c_2}{2}\left({\cal{U}}_2g_{ab}-2{\cal{U}}_1{\cal{K}}_{ab}+2{\cal{K}}^2_{ab}\right)\nonumber\\
&-&\frac{c_3}{2}\left({\cal{U}}_3g_{ab}-3{\cal{U}}_2{\cal{K}}_{ab}+6{\cal{U}}_1{\cal{K}}^2_{ab}-6{\cal{K}}^3_{ab}\right)\nonumber\\
&-&\frac{c_4}{2}\left({\cal{U}}_4g_{ab}-4{\cal{U}}_3{\cal{K}}_{ab}+12{\cal{U}}_2{\cal{K}}^2_{ab}-24{\cal{U}}_1{\cal{K}}^3_{ab}+24{\cal{K}}^4_{ab}\right).
\end{eqnarray}

The first step is to solve the field equations (\ref{4}) and (\ref{5}) for a spherically symmetric geometry which depends on energy. We do that by using the following general line element \cite{3drain, 3dn}
\b
ds^2=-\frac{H(r)}{f^2(\varepsilon)}dt^2+\frac{1}{g^2(\varepsilon)}\left[\frac{dr^2}{H(r)}+r^2d\theta^2\right]. \label{7}\e

The metric function $H(r)$ will be obtained in three different cases of BI, LO, and EX. The rainbow functions $f(\varepsilon)$ and $g(\varepsilon)$represent all the energy dependency of the spacetime, and the coordinates are not energy independent \cite{mag2}. In the limiting case $\varepsilon \rightarrow 0$, the line element (\ref{7}) reduces to that of Einstein-$\Lambda$ gravity theory.

\textbf {In the geometry identified by Eq.(\ref{7}), in terms of a positive constant $c$, the symmetric tensor $f_{a b}$ can be written as \cite{cai}  \b
f_{ab}=\mbox{diag}\left(0,\;0,\;\frac{c^2}{g^2(\varepsilon)}\right).\e Also, in this geometry, the polynomials of Eq.(\ref{2}) take the following explicit forms \cite{3dmpl}}
\b
{\cal{U}}_1=\frac{c}{r},\;\;\;\;\;
\mbox{and}\;\;\;\;\;{\cal{U}}_2={\cal{U}}_3={\cal{U}}_4=0. \label{8}\e

Now, by use of Eqs.(\ref{6}) and (\ref{8}) one can get
\begin{eqnarray}\label{9}
\chi_{tt}&=&-\frac{cc_1}{2r}g_{tt}\nonumber\\
\chi_{rr}&=&\frac{cc_1}{2r}g_{rr},\nonumber\\
\chi_{\theta\theta}&=&0.
\end{eqnarray}
In terms of the non-vanishing component of the electromagnetic tensor $F_{tr}$, we have
\b {\cal{F}}=-2f^2(\varepsilon)g^2(\varepsilon)F^2_{tr}. \label{11}\e

By use of Eqs.(\ref{5}) and (\ref{7}), the electromagnetic field equation may be solved to obtain \cite{dark}

\b F_{tr}(r)=\left\{\begin{array}{ll}\label{12}
	\frac{q}{r\varUpsilon},\;\;(BI)\\\\
	
	\frac{2q}{r(1+\varUpsilon)},\;\;(LO)\\\\
	
	\frac{q}{r}e^{-\frac{L_W}{2}},\;\;(EX)
\end{array} \right.\e

where,
\begin{equation}\label{13}
\varUpsilon=\sqrt{1+\frac{q^2_\varepsilon}{b^2r^2}},
\end{equation}

with

\begin{equation}\label{14}
q_\varepsilon=f(\varepsilon)g(\varepsilon)q,
\end{equation}

where, $q$ is an integration constant, related to the black hole electric charge \cite{dark}. We can see that EX solution obtained in terms of the Lambert function $L_W=Lambert(\xi)$, with $\xi=4(\varUpsilon^2-1)$ \cite{exp}. It is clear that the solution (\ref{12}) reduced to those of charged BTZ black holes as $b\rightarrow\infty$ limit.\\

By using Eq.(\ref{12}) in the relations $F_{tr}=-\partial_r A_t(r)$, we obtain
\b
A_t(r)=-\int F_{tr}(r)dr =\left\{\begin{array}{ll}\label{15}
	-q\ln \left[\frac{r}{\ell}(1+\varUpsilon)\right],\;\;\;\;\;\;\;\;\;\;\;\;\;(BI)\\\\
	
	-q\ln \left[\frac{r}{\ell}(1+\varUpsilon)\right]-\frac{q}{1+\varUpsilon},\;\;\;\;\;\;(LO)\\\\
	
	\frac{q}{2}EI(-\frac{Lw}{2})-qe^{-\frac{Lw}{2}}\;\;\;\;\;\;\;\;\;\;\;\;(EX)
\end{array} \right.\e
for the temporal component of the electromagnetic four-potential. Here, $EI(x)$ is the ExponentIntegral(x) function which is defined as $EI(x)=\int\frac{e^x}{x}dx$. By expanding the values of $F_{tr}$ and $A_t$ in powers of $b^{-1}$, one can show that they reduce to the corresponding values in Maxwell's electrodynamics when the limit $b\rightarrow\infty$ is taken.

In the geometry, introduced by the metric (\ref{7}), the gravitational field
equations (\ref{4}) are reduced to the following differential equations

\b \left.\begin{array}{ll}
C_{tt}=C_{rr}=\frac{H'(r)}{r}+\frac{1}{g^2(\varepsilon)}\left[2\Lambda+4b^2\left(\varUpsilon-1\right)-\frac{cc_1m_G^2}{r}\right]=0,\\\\

C_{\theta\theta}=H''(r)+\frac{1}{g^2(\varepsilon)}\left[2\Lambda+4b^2\left(\varUpsilon^{-1}-1 \right)\right]=0,\end{array} \right\}\;\;\mbox{for}\;\;(BI),\label{16}\e

 \b \left.\begin{array}{ll}
	C_{tt}=C_{rr}=\frac{H'(r)}{r}+\frac{1}{g^2(\varepsilon)}\left\{2\Lambda-8b^2\left[1-\varUpsilon+\ln\left(\frac{1+\varUpsilon}{2} \right) \right]-\frac{cc_1m_G^2}{r}\right\}=0,\\\\
	
	C_{\theta\theta}=H''(r)+\frac{1}{g^2(\varepsilon)}\left[2\Lambda-8b^2\ln\left(\frac{1+\varUpsilon}{2} \right)\right]=0,\end{array} \right\}\;\;\mbox{for}\;\;(LO), \label{17}\e

 \b \left.\begin{array}{ll}
	C_{tt}=C_{rr}=\frac{H'(r)}{r}+\frac{1}{g^2(\varepsilon)}\left[2\Lambda+b^2+\frac{2q_\varepsilon b}{r\sqrt{L_W}}\left(1-L_W\right)-\frac{cc_1m_G^2}{r}\right]=0,\\\\
	
	C_{\theta\theta}=H''(r)+\frac{1}{g^2(\varepsilon)}\left[2\Lambda+b^2-\frac{2q_\varepsilon b}{r\sqrt{L_W}}\right]=0,\end{array} \right\}\;\;\mbox{for}\;\;(EX). \label{18}\e
It is clear that at $g(\varepsilon)=f(\varepsilon)=1$, we recover results of \cite{dark}.\\

Above differential equations satisfy the following relation
\b
\left(r\frac{d}{dr}+1\right)C_{tt}=C_{\theta\theta},\label{19}\e

\begin{figure}[h!]
	\begin{center}$
		\begin{array}{cccc}
		\includegraphics[width=40 mm]{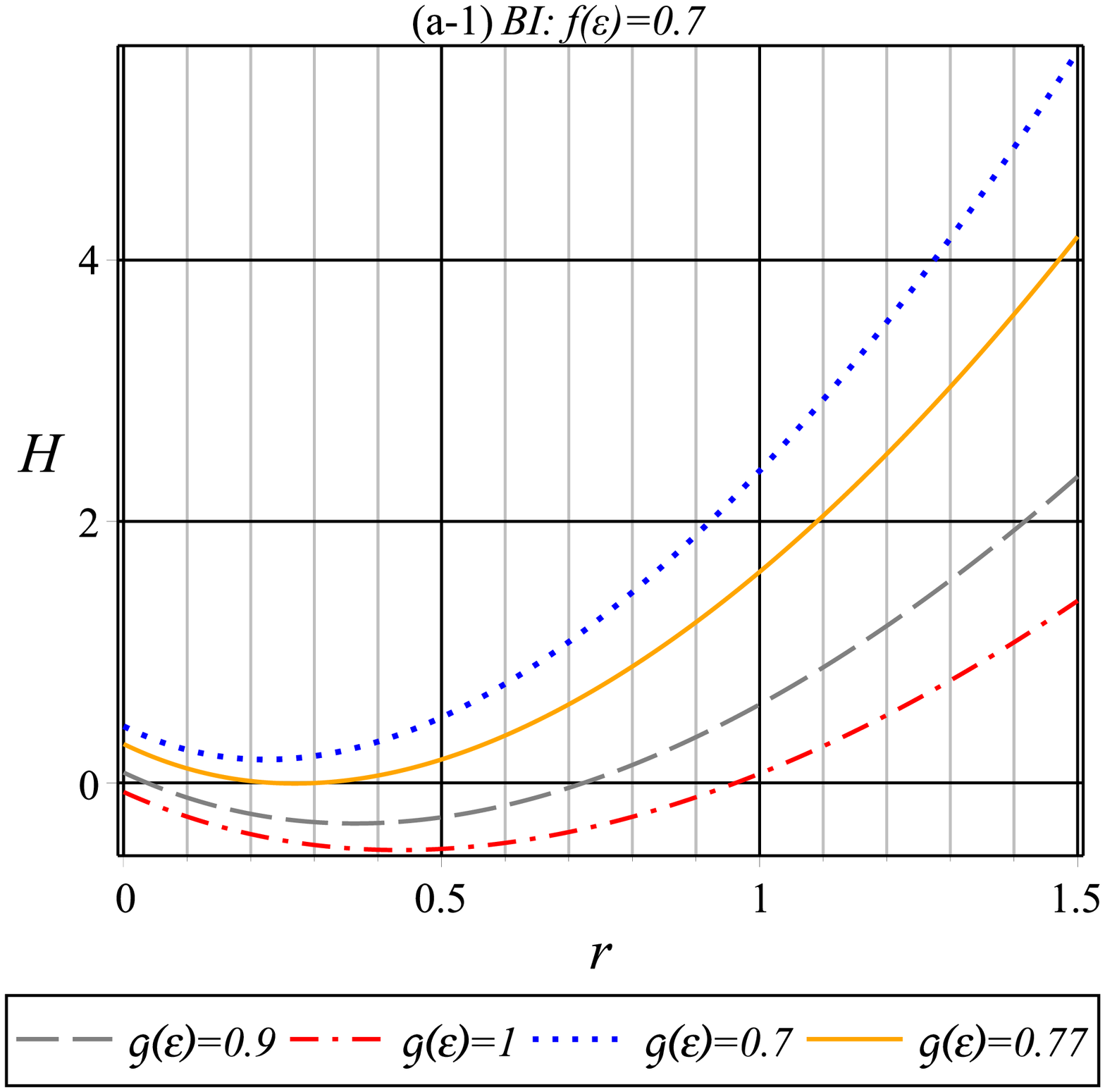}\includegraphics[width=40 mm]{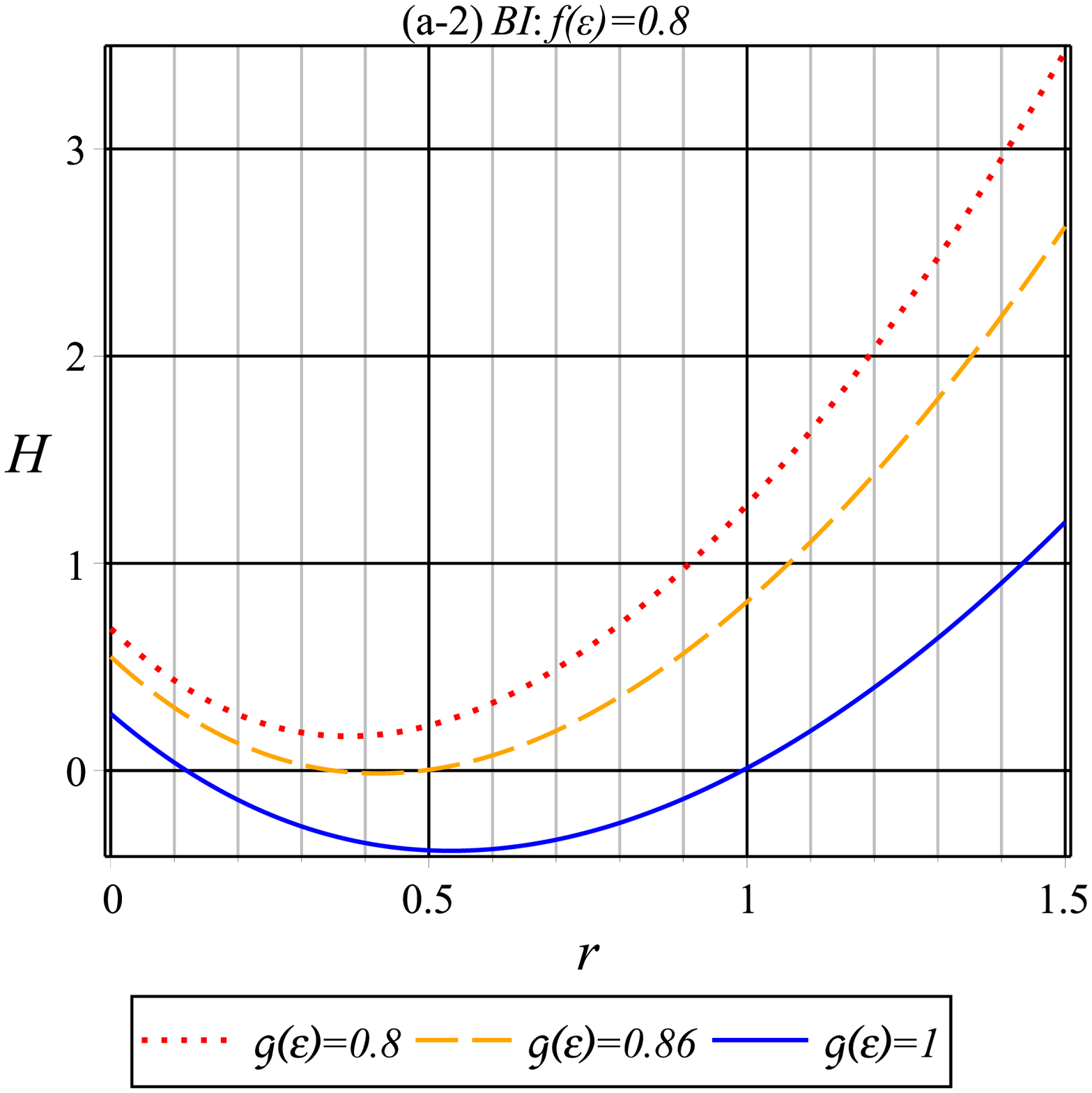}\includegraphics[width=40 mm]{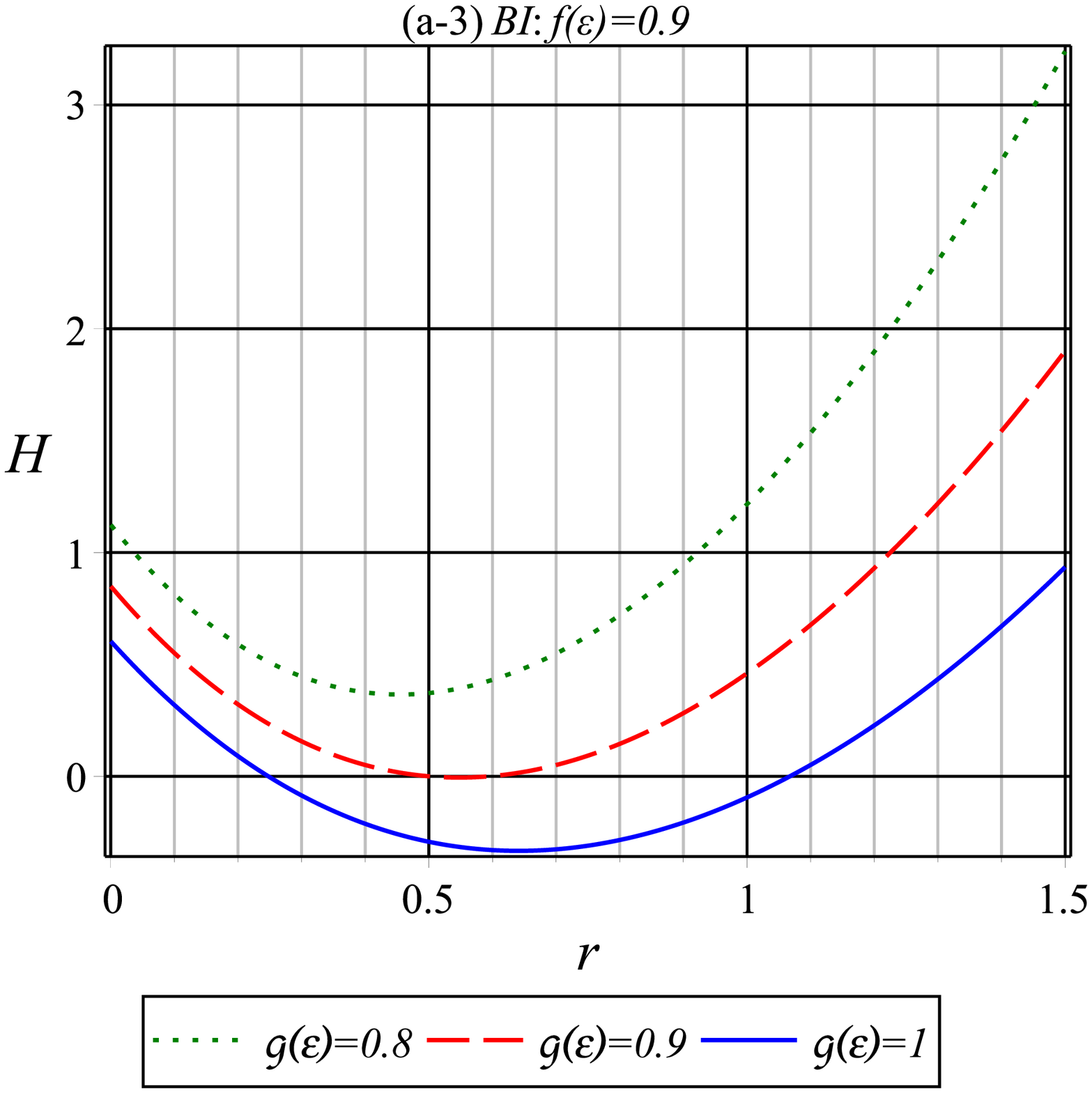}\includegraphics[width=40 mm]{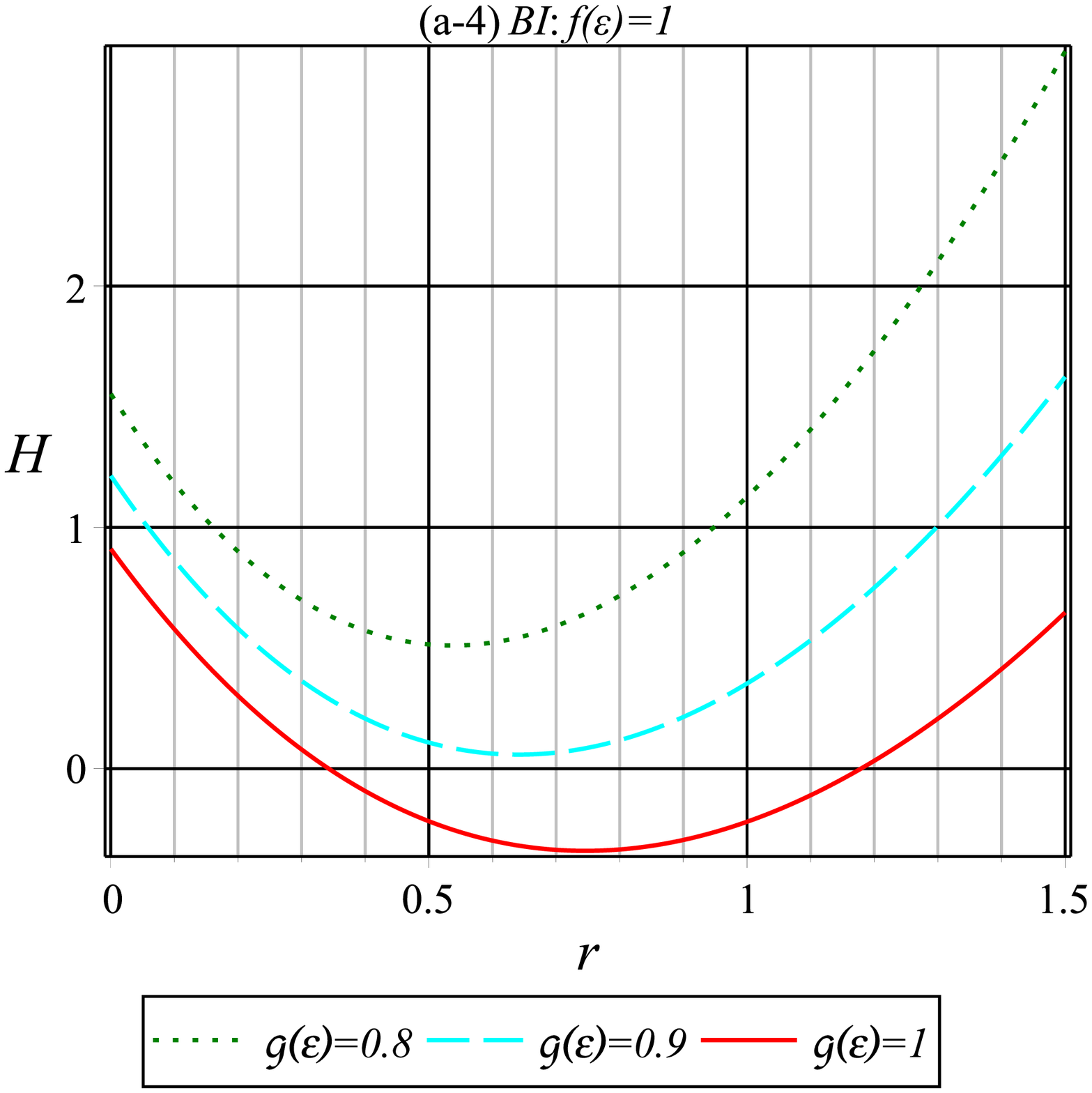}\\
		\includegraphics[width=40 mm]{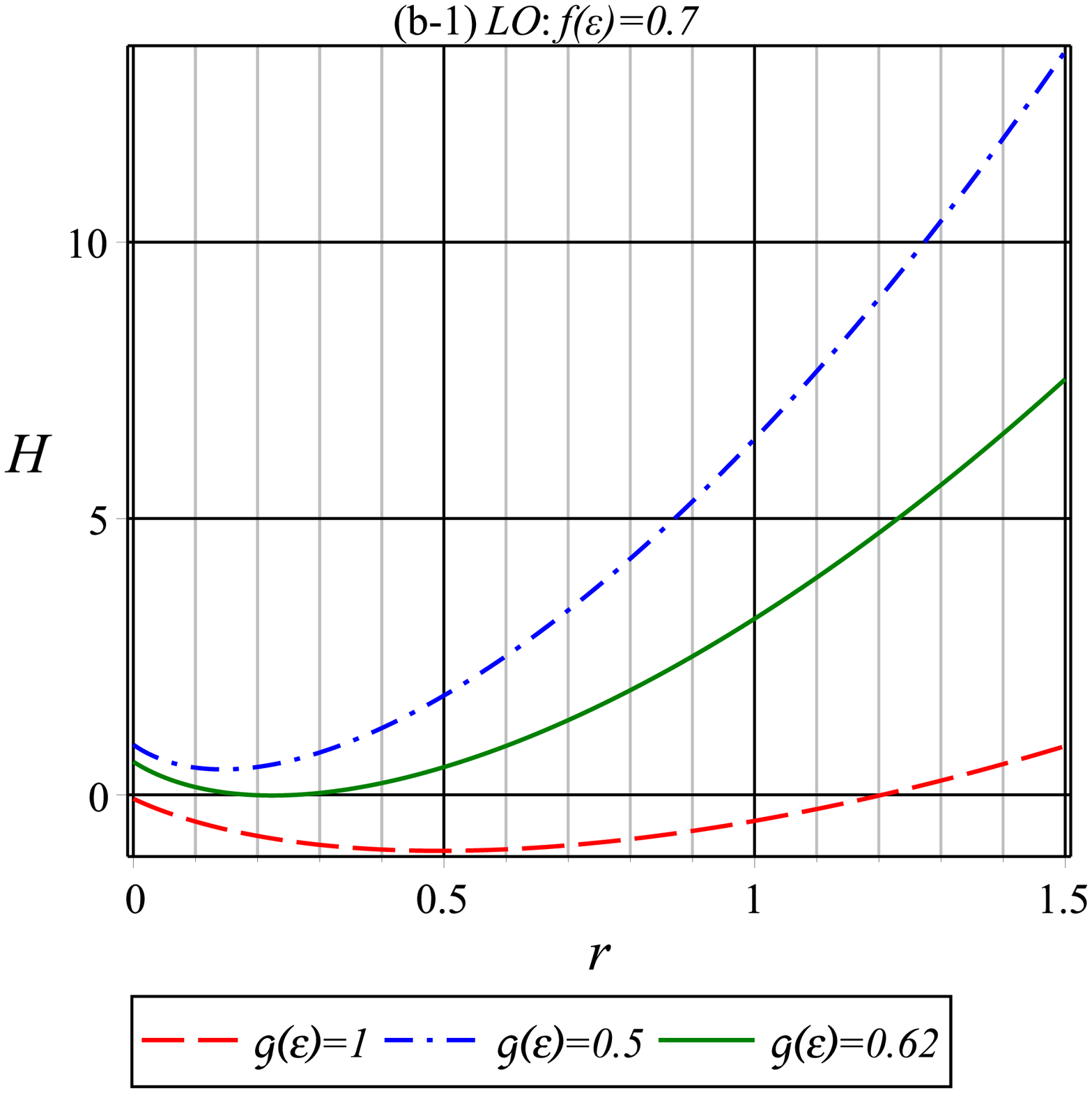}\includegraphics[width=40 mm]{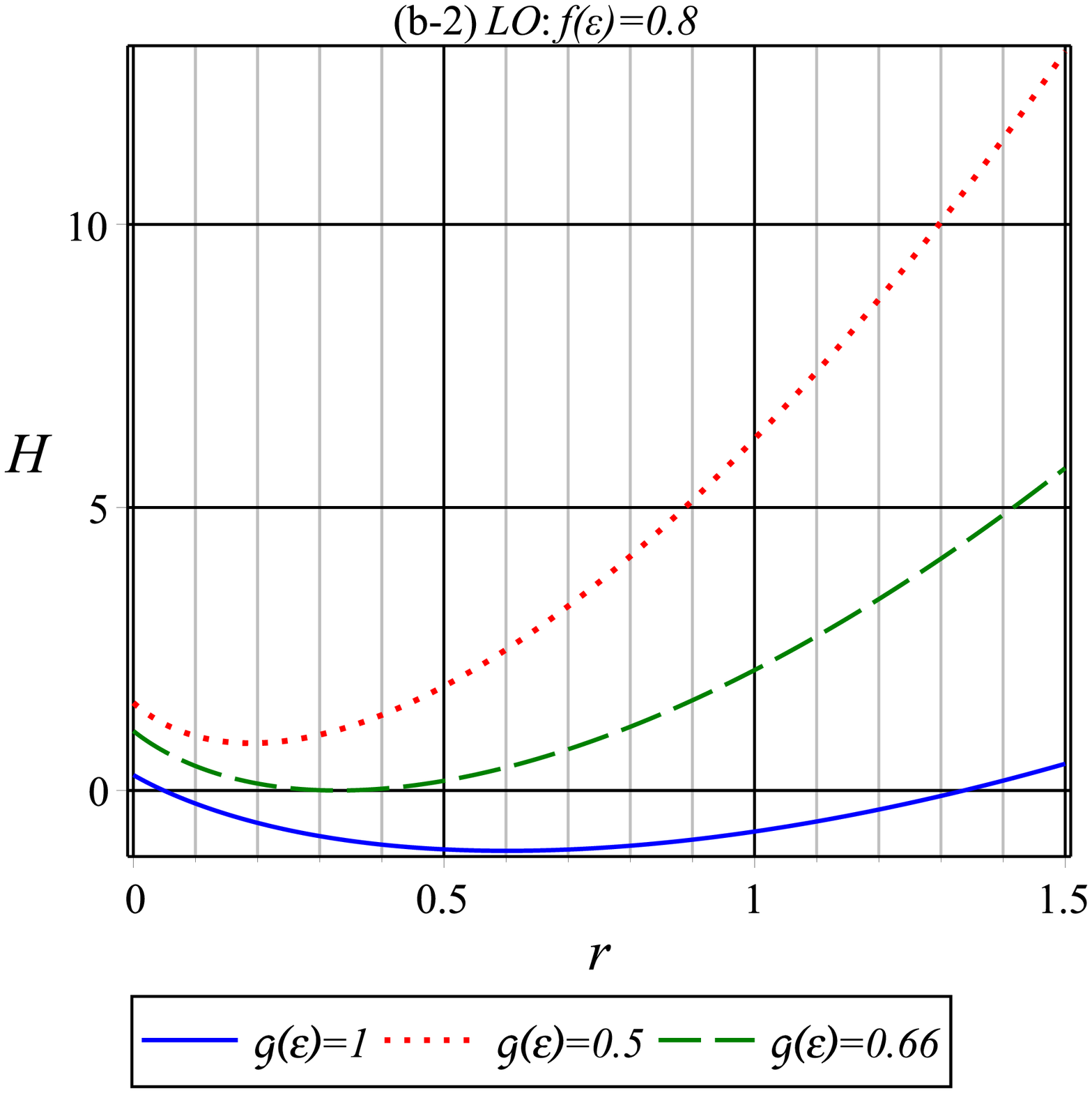}\includegraphics[width=40 mm]{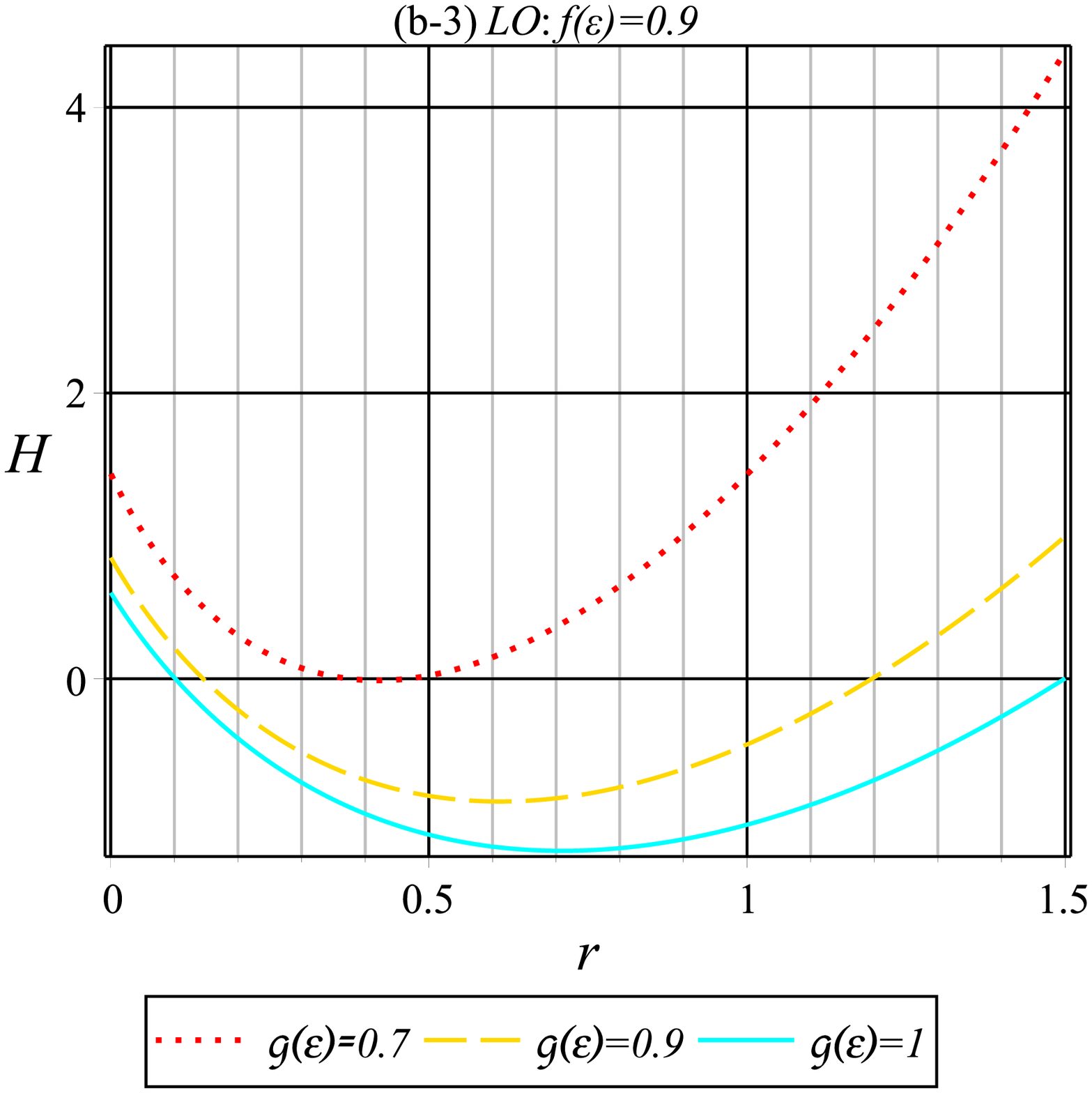}\includegraphics[width=40 mm]{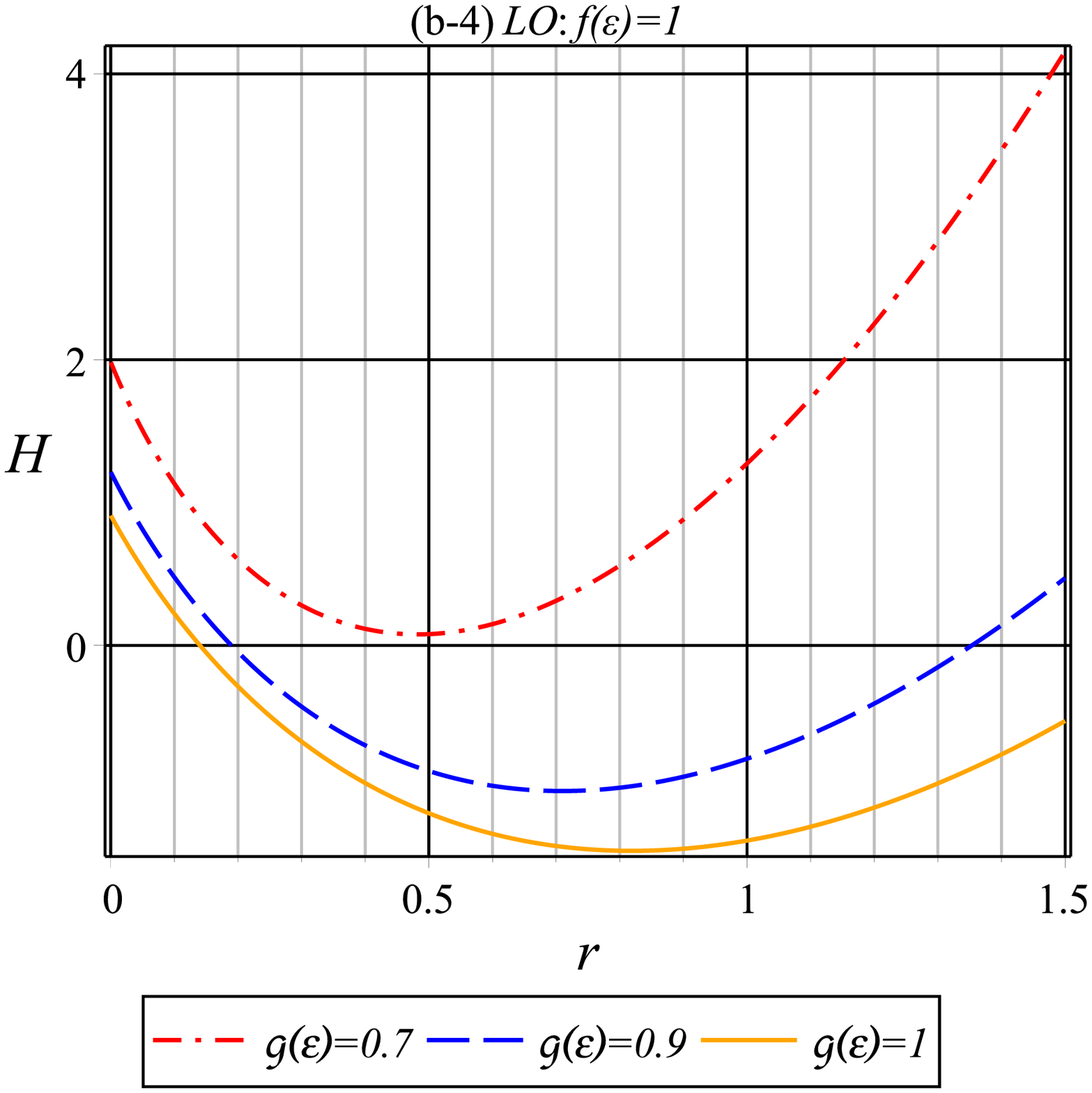}\\
		\includegraphics[width=40 mm]{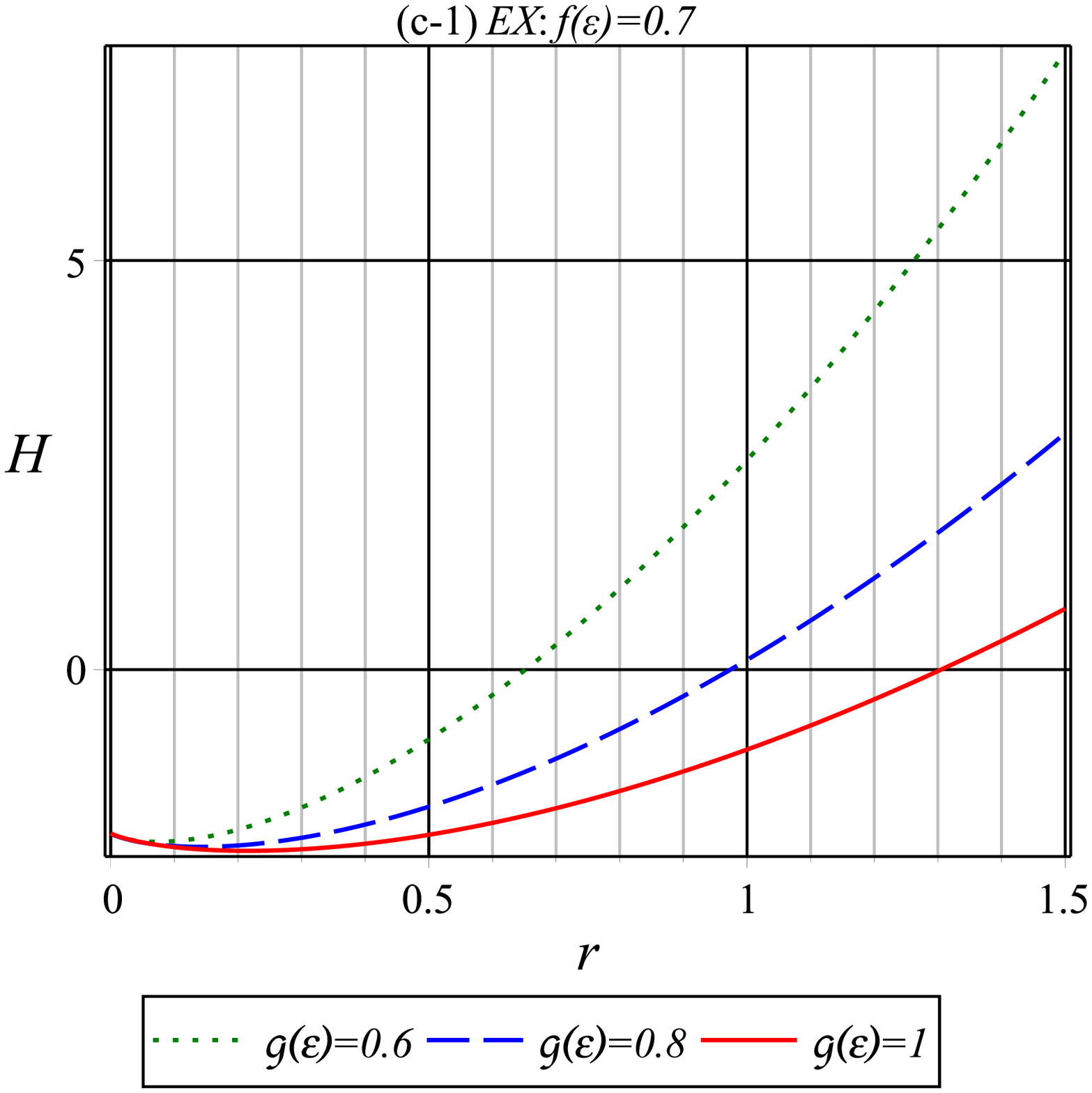}\includegraphics[width=40 mm]{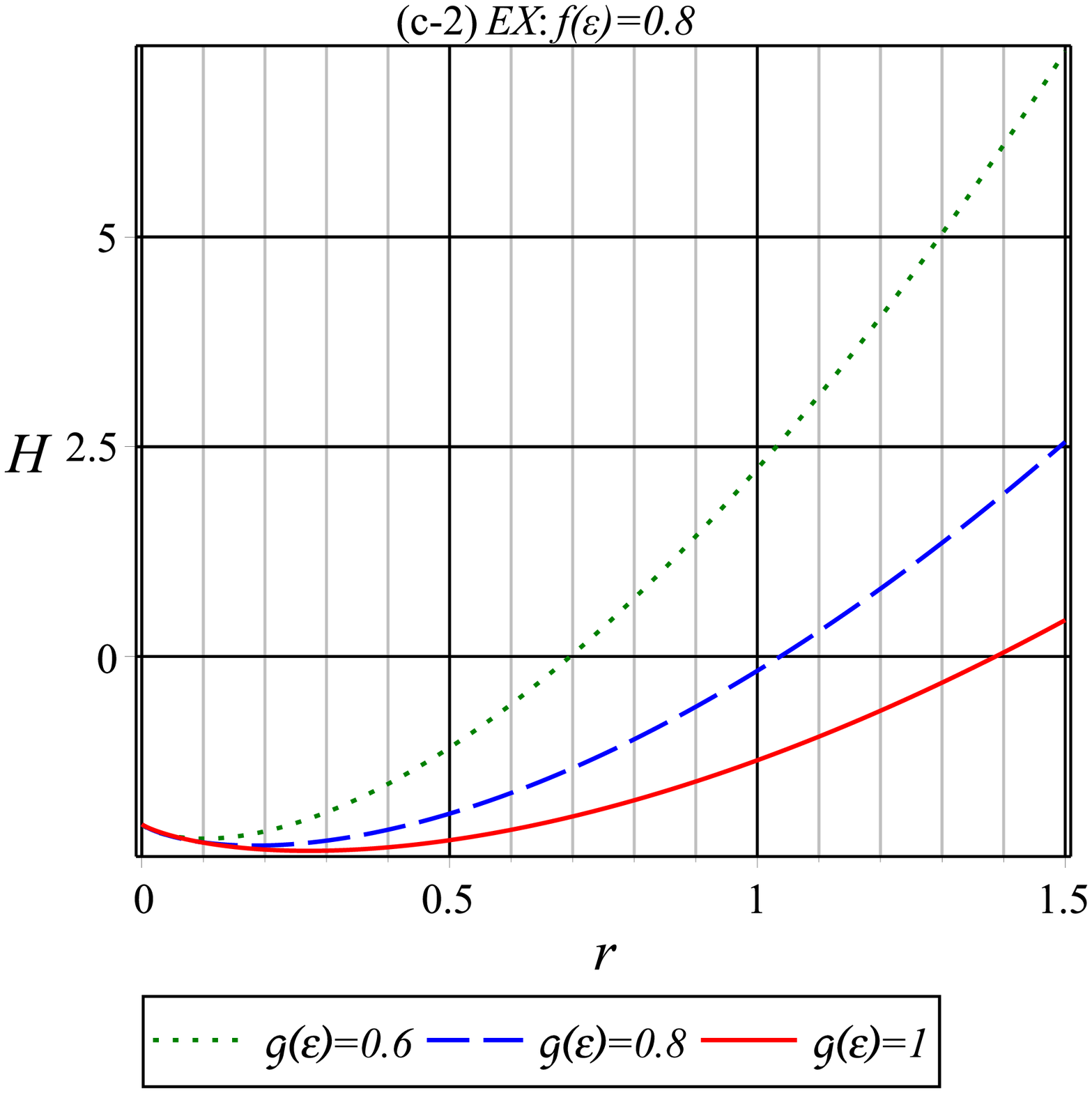}\includegraphics[width=40 mm]{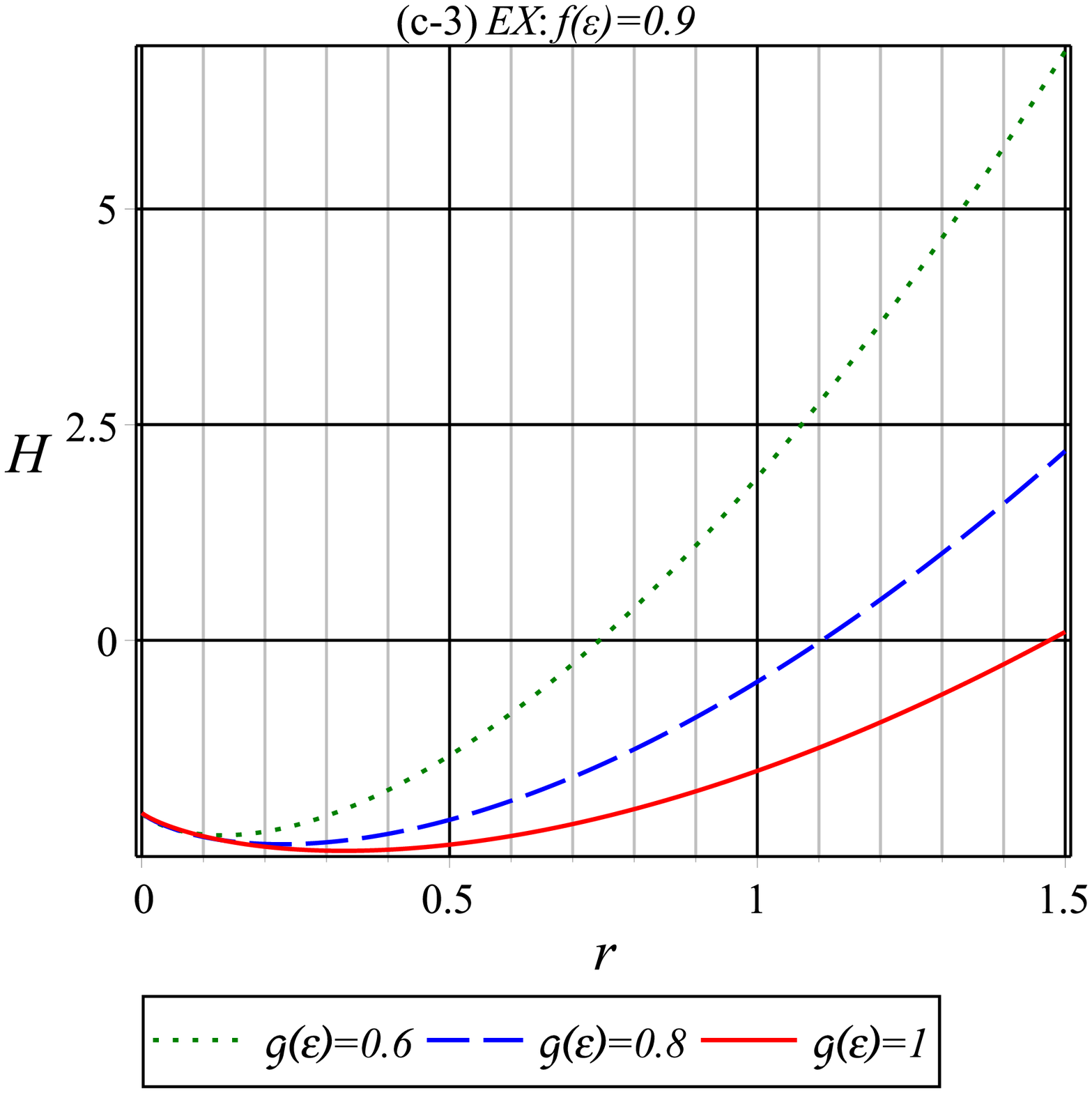}\includegraphics[width=40 mm]{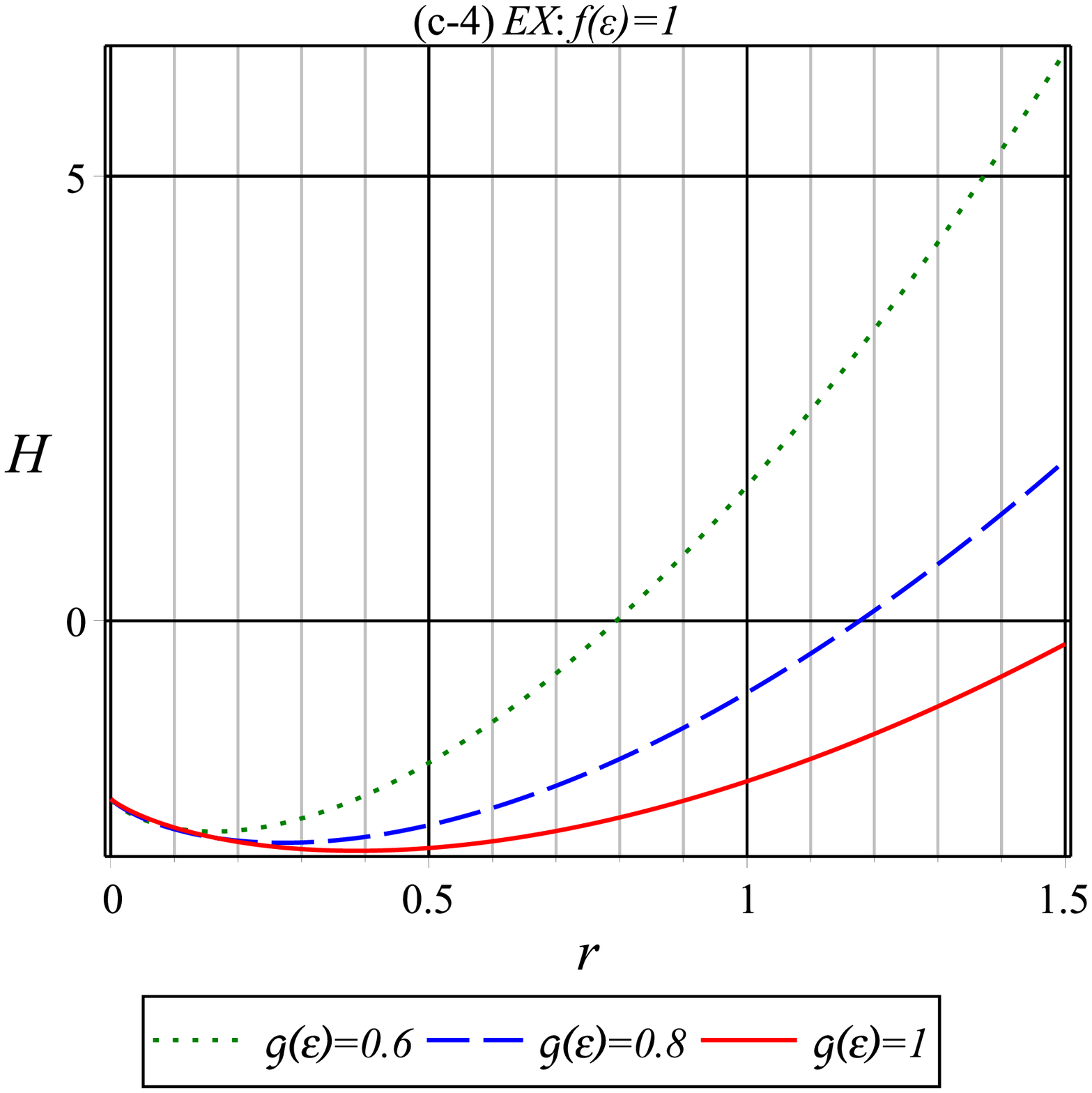}
		\end{array}$
	\end{center}
	\caption{$H(r)$ versus $r$ for $\Lambda=-1$, $c=c_{1}=l=1$. (a) BI with $b=1$, $q=1.2$, $m=2$, $m_G=1.08$ (b) LO with $b=1$, $q=1.2$, $m=2$, $m_G=1.1$ (c) EX with $b=1$, $q=1.2$, $m=2$, $m_G=1$.}
	\label{fig1}
\end{figure}

By use of equation (\ref{19}), we find that two differential equations (\ref{16}) as well as (\ref{17}) and (\ref{18}) are not independent. Hence, we need to solve the first-order differential equations with the fact that the solutions satisfy the second-order differential equation of the same theory.\\
Hence, following Ref. \cite{dark} we can obtain metric function as,

\b
H(r)=\left\{\begin{array}{ll}\label{20}
	-m-\frac{1}{g^2(\varepsilon)}\left(\Lambda r^2-m_G^2cc_1r+H_{BI}\right),\;\;(BI)\\\\
	
	-m-\frac{1}{g^2(\varepsilon)}\left(\Lambda r^2-m_G^2cc_1r+H_{LO}\right),\;\;(LO)\\\\
	
	-m-\frac{1}{g^2(\varepsilon)}\left(\Lambda r^2-m_G^2cc_1r+H_{EX}\right),\;\;(EX)
\end{array} \right.\e
where, $m$ is an integration constant, which is related to the black hole mass. Also, we define

\begin{eqnarray}\label{21}
H_{BI}&=&2b^2r^2\left( \varUpsilon-1\right)-q^2_\varepsilon\left(1-2\ln\left[\frac{r}{2\ell}\left( 1+\varUpsilon\right)\right]\right),\nonumber\\
H_{LO}&=&6b^2r^2\left( \varUpsilon-1\right)-q^2_\varepsilon\left(1-
2 \ln\left[\frac{r}{2\ell}\left( 1+\varUpsilon\right)\right]\right)-4b^2r^2 \ln \left(\frac{1+\varUpsilon}{2}\right),\nonumber\\
H_{EX}&=&\frac{b^2r^2}{2}-q^2_\varepsilon EI\left(-\frac{L_W}{2}\right)-q_\varepsilon br\left(\frac{1}{\sqrt{L_W}}-2\sqrt{L_W} \right).
\end{eqnarray}
It is easy to show that at $b\rightarrow\infty$ limit, where $\varUpsilon=1$, we have $H_{BI}=H_{LO}=H_{EX}\approx2q^2_\varepsilon\ln{(\frac{r}{l})}+ \mbox{constant}$, so we recover massive-BTZ gravity's rainbow \cite{3dmpl}. The effect of $b$ parameter on the horizon structure was already discussed by Ref. \cite{dark}. In Fig. \ref{fig1} we draw $H(r)$ in terms of $r$ to show effect of rainbow variables.\\
The horizon structure shows that, depending on values of $g(\varepsilon)$ and $f(\varepsilon)$, there are regular black hole (with two horizons $r_{\pm}$), black hole with one horizon, extremal case (where $r_{+}=r_{-}$), and naked singularity (where there is no real positive root for $H(r)=0$).\\
Plots of Fig. \ref{fig1} (a), (b) and (c) are corresponding to BI, LO, and EX respectively.
The dash dotted red line of Fig. \ref{fig1} (a-1) shows a black hole with one horizon in BI model, while the dotted red line of Fig. \ref{fig1} (a-2) shows a naked singularity. The extremal black hole has been represented by dashed red line of Fig. \ref{fig1} (a-3) which will be discussed in details by analyzing the black hole temperature in the next section. Also, solid red line of Fig. \ref{fig1} (a-4), which is corresponding to ordinary theory ($f(\varepsilon)=g(\varepsilon)=1$), shows a black hole with two horizons (as well as dotted green lines of Fig. \ref{fig1} (a-1) and (a-3)).\\
Similar results obtained for LO model which is presented by Fig. \ref{fig1} (b-1), (b-2), (b-3) and (b-4). Also, in plots of Fig. \ref{fig1} (c) we can see a black hole with one horizon in EX model.
The space-time singularities can be explained by the curvature scalar. In that case, the Ricci scalar of our model is obtained as
\b
{\cal{R}}=\left\{\begin{array}{ll}\label{22}
	6\Lambda+4b^2\left[\left( \varUpsilon^{-1}-1\right)+2\left(\varUpsilon-1\right)\right]-\frac{2m_G^2cc_1}{r},\;\;(BI)\\\\
	
	6\Lambda+16b^2\left( \varUpsilon-1\right)-24b^2 \ln \left(\frac{1+\varUpsilon}{2}\right)-\frac{2m_G^2cc_1}{r},\;\;(LO)\\\\
	
		6\Lambda+3b^2-\frac{6bq_\varepsilon}{r\sqrt{L_W}}+\frac{4bq_\varepsilon}{r}\sqrt{L_W}-\frac{2m_G^2cc_1}{r},\;\;(EX)
\end{array} \right.\e
which are consistent with those of Einstein-Maxwell-massive black holes in the case $f(\varepsilon)=1=g(\varepsilon)$ \cite{3dmpl}.\\
Also, we can obtain the Kretschmann scalar as following,
\b
{\cal{R}}^{\mu\nu\rho\lambda}{\cal{R}}_{\mu\nu\rho\lambda}=\left\{\begin{array}{ll}\label{23}
12\Lambda^2+16b^4\left[\left( \varUpsilon^{-1}-1\right)^2+2\left(\varUpsilon-1\right)^2\right]-\frac{8b^2 m_G^2cc_1}{r}\left(\varUpsilon-1\right)\\\\\;\;\;\;\;\;\;\;\;\;+16\Lambda b^2\left[\left( \varUpsilon^{-1}-1\right)+2\left(\varUpsilon-1\right)\right]-\frac{8\Lambda m_G^2cc_1}{r}+\frac{2(m_G^2cc_1)^2}{r^2},\;\;(BI)\\\\
	
	12\Lambda^2+64b^4\left\lbrace3 \left[\ln\left(\frac{\varUpsilon+1}{2}\right)\right]^2  +2\left(\varUpsilon-1\right)^2-4\left(\varUpsilon-1\right)\ln\left(\frac{\varUpsilon+1}{2}\right) \right\rbrace \\\\-32\Lambda b^2\left[3\ln\left(\frac{\varUpsilon+1}{2}\right)-2\left(\varUpsilon-1\right)\right]-\frac{8\Lambda m_G^2cc_1}{r}+\frac{2(m_G^2cc_1)^2}{r^2}\\\\\;\;\;\;\;\;\;\;\;\;\;\;\;\;\;\;\;\;\;\;\;\;\;\;\;\;\;\;\;\;\;\;\;\;\;\;\;\;\;\;\;\;+\frac{32b^2m_G^2cc_1}{r}\left[\ln\left(\frac{\varUpsilon+1}{2}\right)-\left(\varUpsilon-1\right)\right],\;\;(LO)\\\\
	
12\Lambda^2+3b^4+\frac{4b^2q^2_\varepsilon}{r^2}\left(\frac{3}{L_W}+2L_W \right) -4(2\Lambda+b^2)\frac{bq_\varepsilon}{r}\left(\frac{3}{\sqrt{L_W}} -2\sqrt{L_W}\right)\\\\ +\frac{2(m_G^2cc_1)^2}{r^2}+\frac{8bq_\varepsilon m_G^2cc_1}{r^2}\left(\frac{1}{\sqrt{L_W}}+\sqrt{L_W} \right)-\frac{16b^2q^2_\varepsilon}{r^2}-\frac{4(2\Lambda+b^2)m_G^2cc_1}{r},\;(EX)
\end{array} \right.\e
Now, from Eqs.(\ref{22}) and (\ref{22}), one can find that the Ricci and Kretschmann scalars are finite for finite values of $r$. There is an essential (not coordinate) singularity located at the origin, and the asymptotic behavior of our solutions is just like the AdS black holes \cite{dark}.

\setcounter{equation}{0}
\section{Thermodynamic properties}
In the previous section, we obtained new black hole solutions corresponding to three different models. Now, we study thermodynamics of these black holes.
The temperature of BI, LO, and EX solutions are given by the following relations
\b T=\left\{\begin{array}{ll}\label{24}
	\frac{1}{4\pi f(\varepsilon) g(\varepsilon)}\left[m_G^2cc_1-2\Lambda
	r_+-\frac{4q^2_{\varepsilon}}{r_+(1+\varUpsilon_+)}\right],\;\;(BI)\\\\
	
	\frac{1}{4\pi f(\varepsilon) g(\varepsilon)}\left\{m_G^2cc_1-2\Lambda r_+
	 -8b^2 r_+\left[\varUpsilon_+-1-\ln \left(  \frac{1+\varUpsilon_+}{2}\right) \right] \right\},\;\;(LO)\\\\
	
\frac{1}{4\pi f(\varepsilon) g(\varepsilon)}\left[m_G^2cc_1-(2\Lambda+b^2)r_+ +2bq_\varepsilon\left(\frac{1}{\sqrt{L_{W+}}}-\sqrt{L_{W+}} \right)\right],\;\;(EX)
\end{array} \right.\e
in which $\varUpsilon_+=\sqrt{1+\frac{q^2_{\varepsilon}}{b^2r^2_+}}$, also $L_{W+}=L_{W}(\xi_+)$ with $\xi_+=4(\varUpsilon^2_+-1)$.

We can also obtain the condition of extreme black holes, which is $T(r_{ext},q_{ext})=0$. It leads to the following equations
\b 2\Lambda r_{ext}+\frac{4q^2_{\varepsilon\;ext}}{r_{ext}(1+\varUpsilon_{ext})}=m_G^2cc_1,\;\;(BI),\label{25}\e

\b 2\Lambda r_{ext}
+8b^2 r_{ext}\left[\varUpsilon_{ext}-1-\ln \left(\frac{1+\varUpsilon_{ext}}{2}\right) \right]=m_G^2cc_1,\;\;(LO),\label{26}\e

\begin{figure}[h!]
 \begin{center}$
 \begin{array}{cccc}
\includegraphics[width=55 mm]{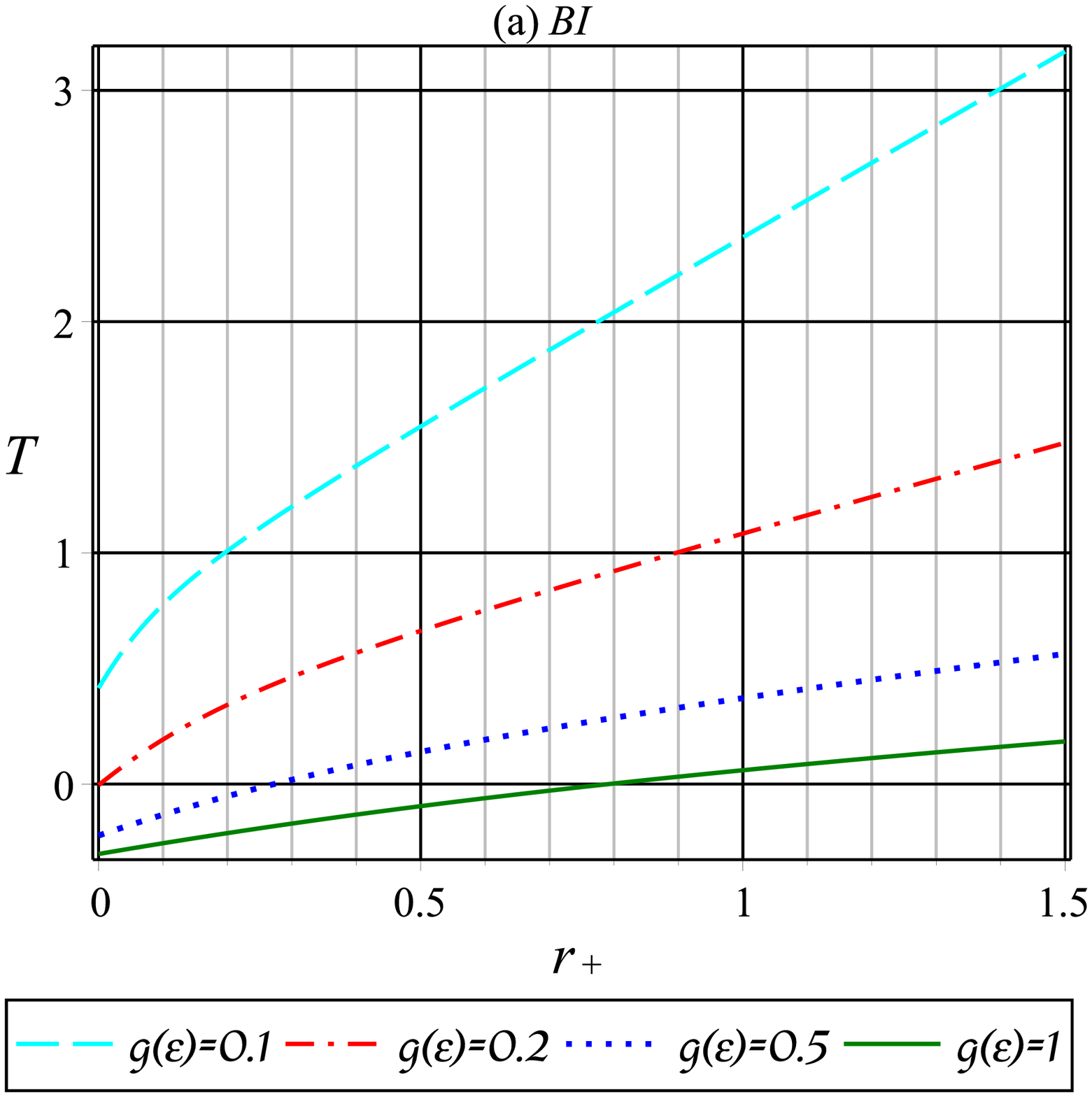}
\includegraphics[width=55 mm]{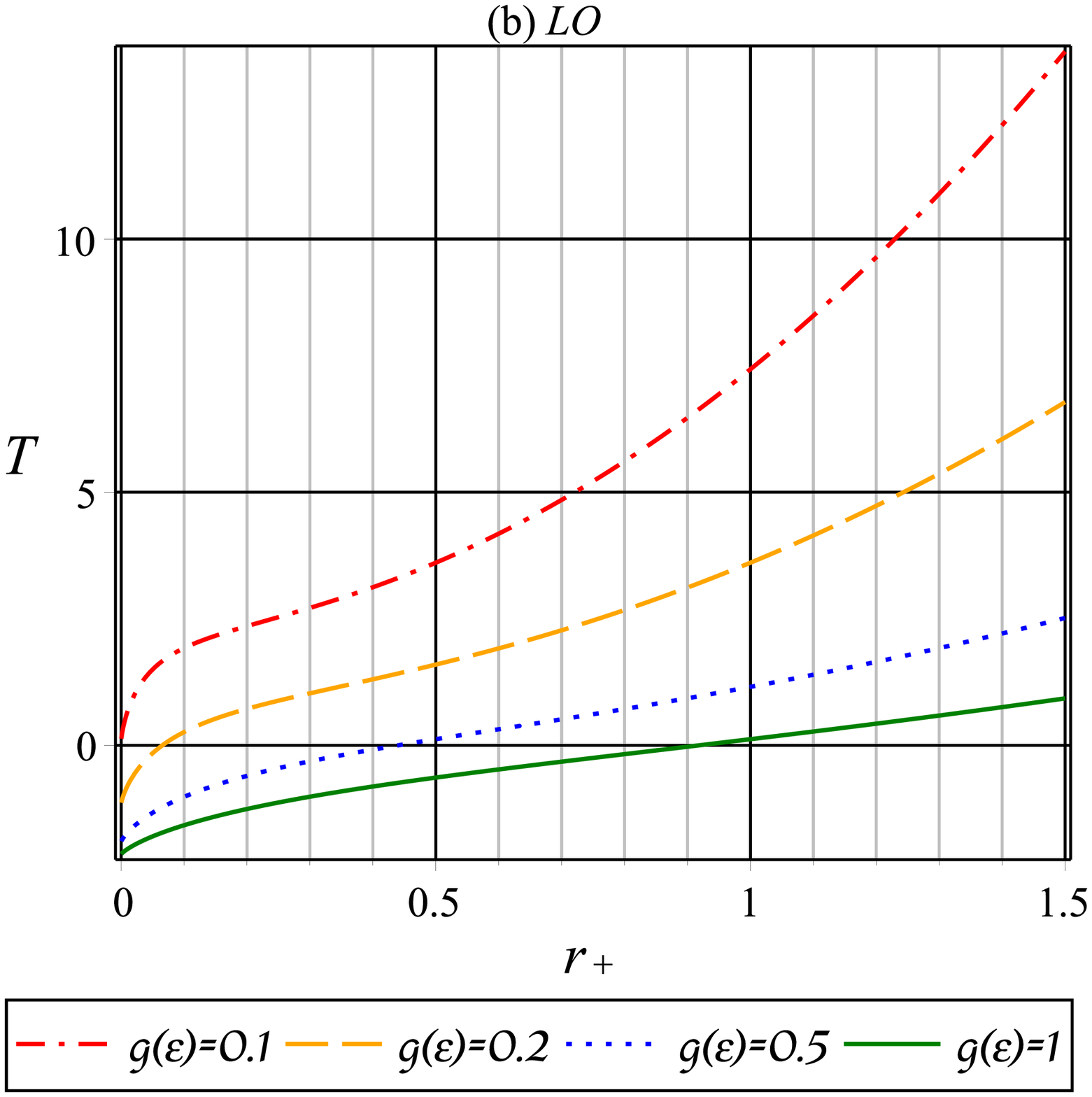}
\includegraphics[width=55 mm]{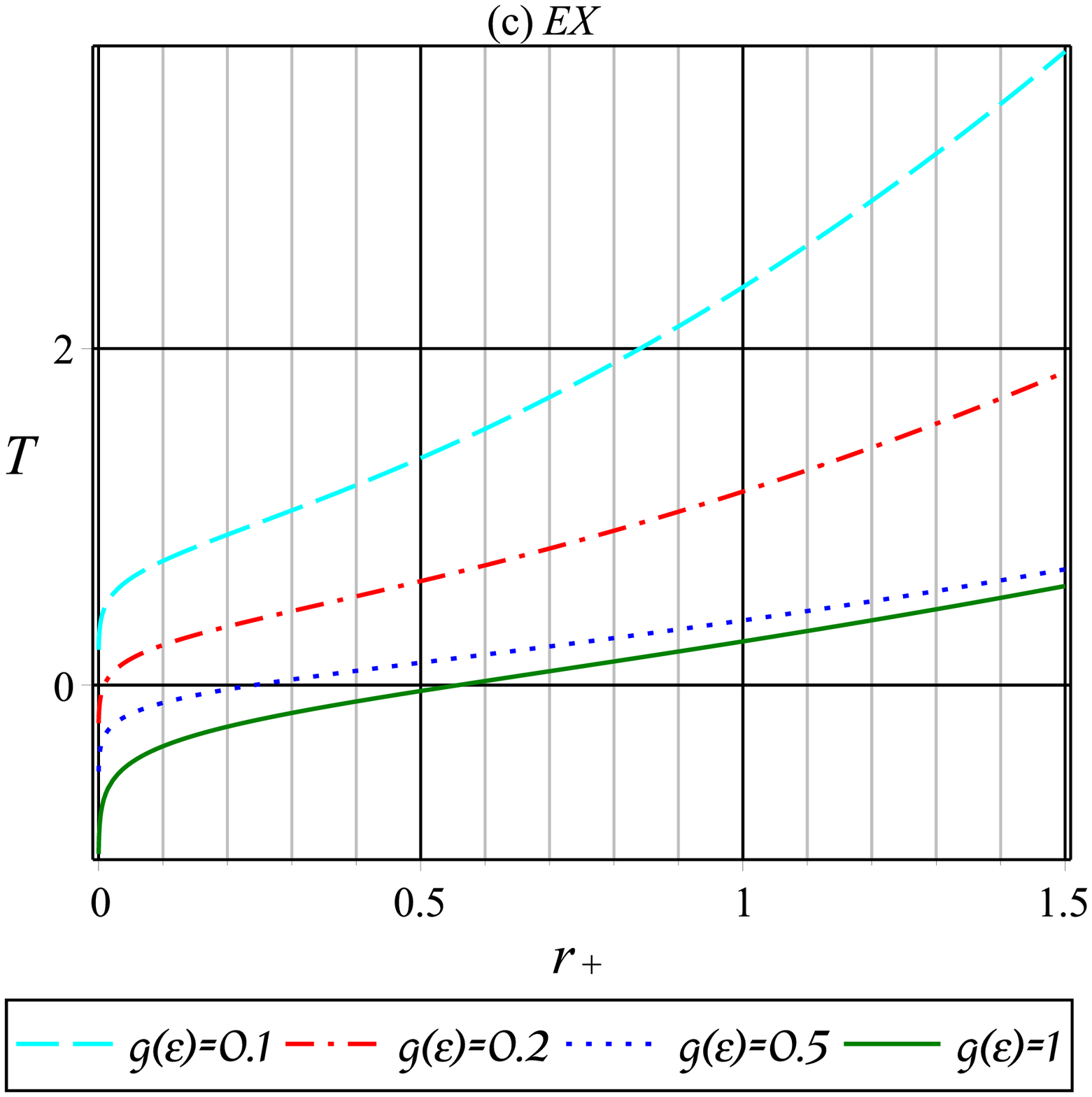}
 \end{array}$
 \end{center}
\caption{Temperature versus $r_{+}$ for $\Lambda=-1$, $c=c_{1}=l=1$, $b=1$, $q=1.2$, $m=2$, $m_G=1$ and $f(\varepsilon)=1$. (a) BI; (b) LO; (c) EX.}
 \label{fig2}
\end{figure}

\b (2\Lambda+b^2)r_{ext} +2bq_{\varepsilon\;ext}\left(\sqrt{L_{W{ext}}}-\frac{1}{\sqrt{L_{W{ext}}}} \right)=m_G^2cc_1,\;\;(EX).\label{27}\e

where, we defined $q_{\varepsilon\;ext}=f(\varepsilon)g(\varepsilon)q_{ext}$,  $\varUpsilon_{ext}=\sqrt{1+\frac{q^2_{\varepsilon\;ext}}{b^2r^2_{ext}}}$ and $L_{W{ext}}=L_{W}(\xi_{ext})$ with $\xi_{ext}=4(\varUpsilon^2_{ext}-1)$.
In order to find more information about extremal black holes we draw temperature given by equation (\ref{24}) in terms of $r_{+}$. In Fig. \ref{fig2} we can see that the black hole temperature is positive for $r_{+}>r_{ext}$. Plots of Fig. \ref{fig2} (a), (b) and (c) are corresponding to BI, LO, and EX respectively. Generally, it is increasing function of horizon radius. For example, from Fig. \ref{fig2} (a) we can see that $r_{ext}=0.8$ in BI model for $f(\varepsilon)=g(\varepsilon)=1$ (see solid green line). Also, for the small values of $g(\varepsilon)$, we can see that there is no extremal case (for example see cyan dashed line of Fig. \ref{fig2} (a)).

The Hawking-Bekenstein entropy of nonlinearly charged BTZ black holes in massive gravity's rainbow is given by
\b
S_{0}=\frac{\pi r_+}{2g(\varepsilon)}.\label{28}\e Here, we have used the system of units with $l_{p}=1$. Appearance of the rainbow function $g(\varepsilon)$, shows that the black hole entropy is affected by gravity's rainbow. In the case of $ g(\varepsilon)=1$, we recover the results of nonlinearly charged BTZ black holes in massive gravity \cite{dark}. It should be noted that the entropy given by (\ref{28}) coincides with Wald's entropy \cite{01, 02, 03}.\\
The electric potential of this model is given by \cite{mh, 2017, 20172},

\b \Phi(r_+)=\left\{\begin{array}{ll}\label{29}
	-q\ln\left[\frac{r_+}{2\ell}\left( 1+\varUpsilon_+\right)\right],\;\;(BI)\\\\
	
	-q\ln\left[\frac{r_+}{2\ell}\left( 1+\varUpsilon_+\right)\right]-\frac{q}{1+\varUpsilon_+},\;\;(LO)\\\\
	
	\frac{q}{2}EI(-\frac{L_{W+}}{2})-qe^{-\frac{L_{W+}}{2}},\;\;(EX)
\end{array} \right.\e

One can calculate the conserved electric charge $Q$ by using Gauss's electric law \cite{20172},
\b Q=\frac{f(\varepsilon)}{2}q_{\varepsilon}. \label{30}\e

Also, the black hole mass $M$ is given by\cite{3drain},

\b  M=\frac{m}{8f(\varepsilon)} \label{31}\e

The mass parameter $m$ is related to the horizon radius via $H(r_+)=0$. These relations show that the black hole mass and charge get modified in the presence of rainbow functions. In the low energy regime (i.e. $f(\varepsilon)=g(\varepsilon)=1$) these quantities reduce to those of BTZ black holes.\\

We can check validity of the first law of black hole thermodynamics using the conserved and thermodynamic quantities presented in equations (\ref{24}), (\ref{28}), (\ref{29}), (\ref{30}) and (\ref{31}). To do that, we need to calculate a Smarr-type mass formula which is given by \b
M(Q,S_{0})=\left\{\begin{array}{ll}\label{32}
	\frac{-S_{0}^2}{4f(\varepsilon)g(\varepsilon)}\left\{\frac{2\Lambda g(\varepsilon)}{\pi^2}-\frac{m_G^2cc_1}{\pi S_{0}}+M_{BI}\right\},(BI)\\\\
	
\frac{-S_{0}^2}{4f(\varepsilon)g(\varepsilon)}\left\{\frac{2\Lambda g(\varepsilon)}{\pi^2}  -\frac{m_G^2cc_1}{\pi S_{0}}+M_{LO} \right\},\;\;(LO)\\\\
	
\frac{-S_{0}^2}{4f(\varepsilon)g(\varepsilon)}\left\{\frac{2\Lambda g(\varepsilon)}{\pi^2} -\frac{m_G^2cc_1}{\pi S_{0}}+M_{EX} \right\},\;\;(EX)
\end{array} \right.\e

where, we defined
\begin{eqnarray}\label{33}
M_{BI} &=& \frac{4b^2g(\varepsilon)}{\pi^2}\left( \varUpsilon_{S_{0}}-1\right)-\frac{2g(\varepsilon)Q^2}{S_{0}^2}\left( 1- \ln\left[\frac{g(\varepsilon) S_{0}}{\pi\ell}\left( 1+\varUpsilon_{S_{0}}\right)\right]\right) ,\nonumber\\
M_{LO} &=& \frac{12b^2g(\varepsilon)}{\pi^2}\left( \varUpsilon_{S_{0}}-1\right)-\frac{8b^2g(\varepsilon)}{\pi^2}\ln\left(\frac{ \varUpsilon_{S_{0}}+1}{2}\right)-\frac{2g(\varepsilon)Q^2}{S_{0}^2}\left( 1- \ln\left[\frac{g(\varepsilon) S_{0}}{\pi\ell}\left( 1+\varUpsilon_{S_{0}}\right)\right]\right)\nonumber\\
M_{EX} &=& \frac{b^2g(\varepsilon)}{\pi^2}-\frac{2g(\varepsilon)Q^2}{S_{0}^2}EI(-\frac{L_{WS_{0}}}{2})-\frac{2bg(\varepsilon)}{\pi S_{0}}Q\left(\frac{1}{\sqrt{L_{WS_{0}}}}-2\sqrt{L_{WS_{0}}}\right),
\end{eqnarray}
where, $\varUpsilon_{S_{0}}=\sqrt{1+\frac{\pi^2Q^2}{b^{2}S_{0}^2}}$, while $L_{WS_{0}}=L_{W}(\xi_{S_{0}})$, with $\xi_{S_{0}}=\frac{4\pi^2Q^2}{b^{2}S_{0}^2}$. In that case, we have
\begin{eqnarray}\label{34}
\Phi &=& \left(\frac{\partial M }{\partial Q}\right)_{S_{0}},\nonumber\\
T &=& \left(\frac{\partial M}{\partial S_{0}} \right)_Q.
\end{eqnarray}

Hence, the first law  of black hole thermodynamics is valid in its standard form. That is
\b dM=TdS_{0}+\Phi dQ, \label{35}\e

Therefore, one can say that, although almost all of the conserved and thermodynamic quantities get modified due to the consideration of rainbow functions, they satisfy the standard form of the thermodynamical first law.\\

With the thermodynamic quantities in hand, we can investigate thermodynamic stability or phase transitions as well as the critical points. Local stability of the black holes, in the canonical ensemble, is studied by analyzing the signature  of specific heat. The specific heat is positive for the black hole which is thermodynamically stable. On the other hand, the black hole with negative specific heat is unstable. In that case,  the first or the second-order phase transition is possible.\\

A first-order phase transition point is determined by the vanishing specific heat. On the other hand, the divergent point of specific heat shows the second order phase transition. The black hole specific heat obtained using the following formula \cite{3dst, ijmpd, setare},
\b {\cal{C}}_Q =T\left(\frac{\partial S_{0}}{\partial T}\right)_Q.\label{36}\e

Hence, one can obtain,

\b{\cal{C}}_Q=\left\{\begin{array}{ll}\label{37}
	\frac{\pi \varUpsilon_+\left[m^2_Gcc_1-4b^2r_+(\varUpsilon_+-1)-2\Lambda r_+ \right]} {4g(\varepsilon)\left[2b^2(\varUpsilon_+-1)-\Lambda \varUpsilon_+ \right]},\;\;(BI)\\\\
	
\frac{\pi \left[m^2_Gcc_1-8b^2r_+\left\lbrace \varUpsilon_+-1-\ln\left(\frac{1+\varUpsilon_+}{2} \right) \right\rbrace -2\Lambda r_+ \right]} {4g(\varepsilon)\left[4b^2\ln\left(\frac{1+\varUpsilon_+}{2}\right)-\Lambda \right]},\;\;(LO)\\\\
	
	\frac{\pi\left[m_G^2cc_1-(2\Lambda+b)r_+ +2bq_\varepsilon\left(\frac{1}{\sqrt{L_{W+}}}-\sqrt{L_{W+}} \right)\right]} {2g(\varepsilon)\left[b^2(e^{\frac{L_{W+}}{2}}-1)-2\Lambda  \right]},\;\;(EX)
\end{array} \right.\e

\begin{figure}[h!]
 \begin{center}$
 \begin{array}{cccc}
\includegraphics[width=55 mm]{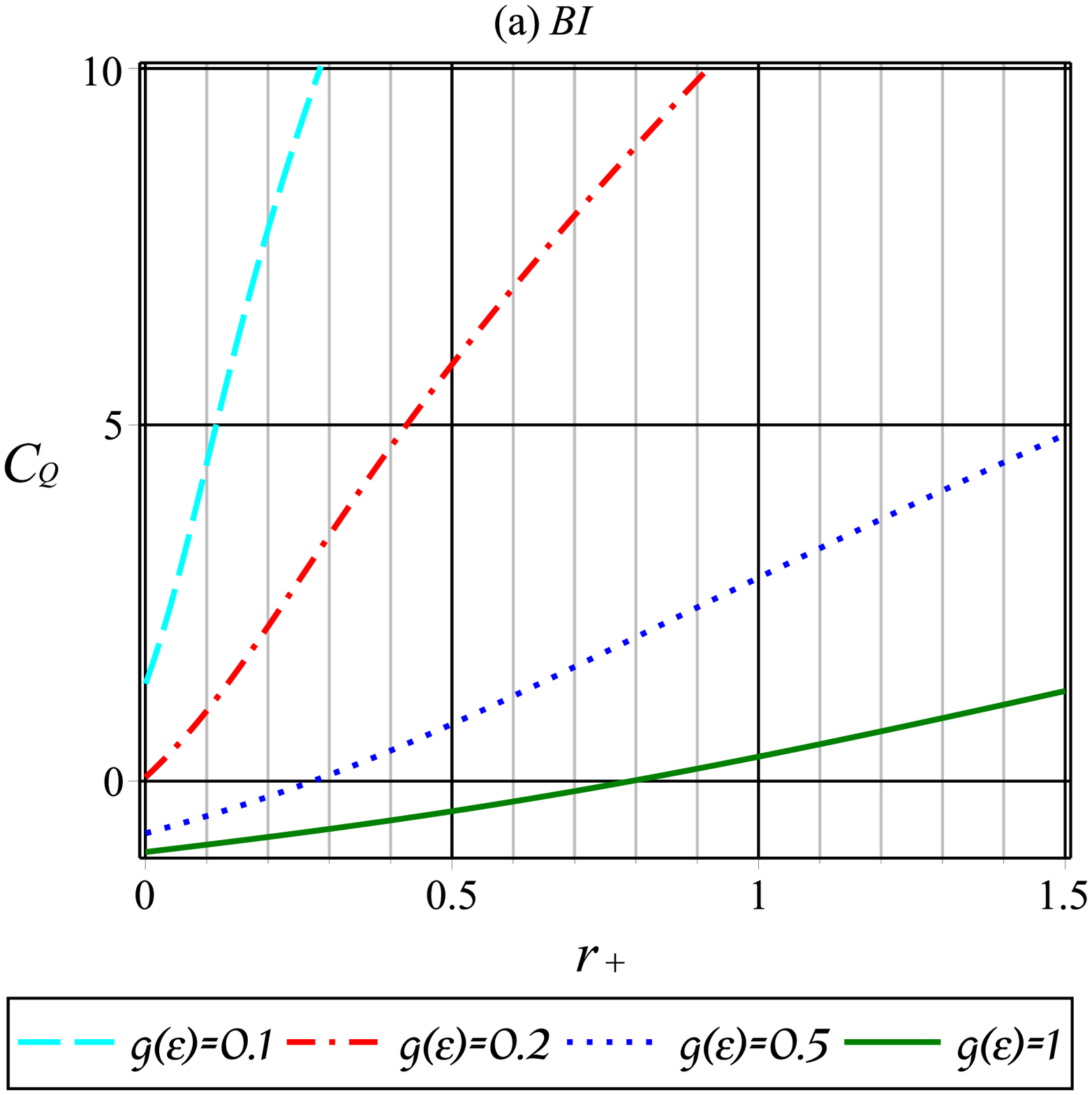}
\includegraphics[width=55 mm]{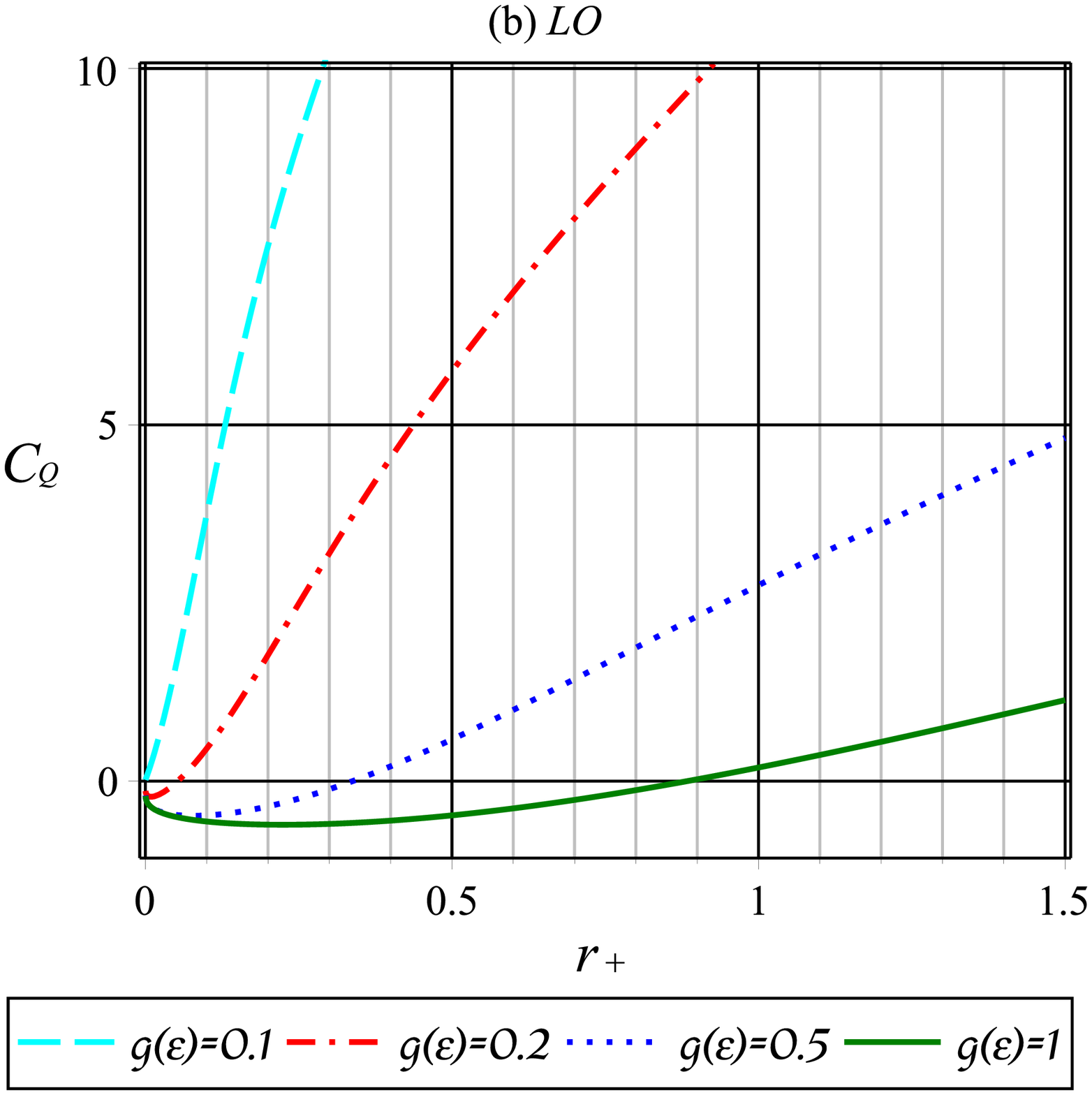}
\includegraphics[width=55 mm]{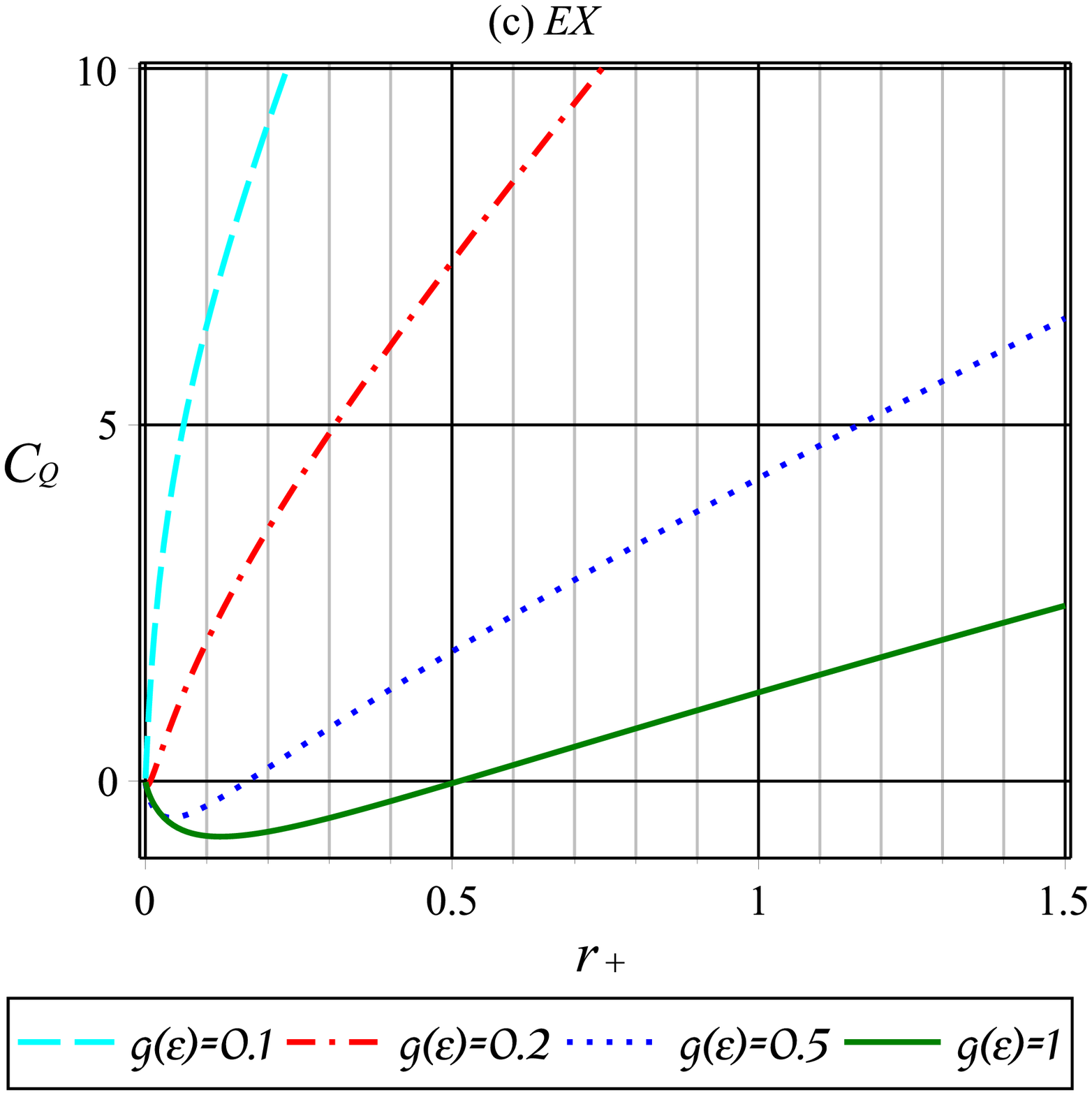}
\end{array}$
\end{center}
\caption{Specific heat versus $r_{+}$ for $\Lambda=-1$, $c=c_{1}=l=1$, $b=1$, $q=1.2$, $m=2$, $m_G=1$ and $f(\varepsilon)=1$. (a) BI; (b) LO; (c) EX.}
 \label{fig3}
\end{figure}

In plots of Fig. \ref{fig3} we draw specific heat in terms of horizon radius and verify the first-order phase transition.

Already we found that (see Fig. \ref{fig2}) our black hole temperature is positive for $r_{+}\geq r_{ext}$. Now, from plots of Fig. \ref{fig3} we can see that $C_{Q}\geq0$ for $r_{+}\geq r_{ext}$. It means that a black hole with positive temperature is completely stable, and $r_{+}=r_{ext}$ is minimum value for the event horizon radius where the black hole can exist. It occur for three cases of BI, LO and EX which are presented by Fig. \ref{fig3} (a), (b) and (c) respectively.\\
For the small values of $g(\varepsilon)$ we can see that $r_{ext}=0$ and the black hole is completely stable for any values of horizon radius. However, there is no second-order phase transition, as there is no asymptotic behavior.\\

%\newpage
\setcounter{equation}{0}
\section{Corrected thermodynamics}
Now, we would like to use the special case of the entropy (\ref{s}), as
\begin{equation}\label{sLn}
S=S_{0}+\alpha \ln{S_{0}},
\end{equation}
which is the logarithmic corrected entropy, also, the entropy $S_{0}$ is given by Eq.(\ref{28}).
We study the modified thermodynamics due to the logarithmic correction for three different models of BI, LO, and EX. Quantum corrections can be considered as the small perturbations around the equilibrium. Consideration of logarithmic corrected entropy leads to the modification of the first law. Through some algebraic calculations, we obtain
\begin{equation}\label{first law BI}
dM=TdS+\Phi dQ-\alpha T d\lambda,
\end{equation}
where,
\begin{equation}\label{A BI}
d\lambda=\frac{dS_{0}}{S_{0}},
\end{equation} and \begin{equation}
\lambda=\ln S_{0}.
\end{equation}
Indeed we introduced $\lambda$ as a new thermodynamic parameter responsible for logarithmic correction term, and coefficient $\alpha T$ as its conjugate variable. Hence, the thermodynamic relations (\ref{34}) must be extended as

\begin{eqnarray}\label{34-Ln}
\Phi &=& \left(\frac{\partial M }{\partial Q}\right)_{S, \lambda},\nonumber\\
T &=& \left(\frac{\partial M}{\partial S} \right)_{Q, \lambda},\nonumber\\
\alpha &=& -\frac{1}{T}\left(\frac{\partial M}{\partial \lambda} \right)_{Q, S}.
\end{eqnarray}

It may help us to fix the thermal fluctuation coefficient for various horizon radius. Now, we can study modified thermodynamics for three different models.
\subsection{BI}
\begin{figure}[h!]
 \begin{center}$
 \begin{array}{cccc}
\includegraphics[width=65 mm]{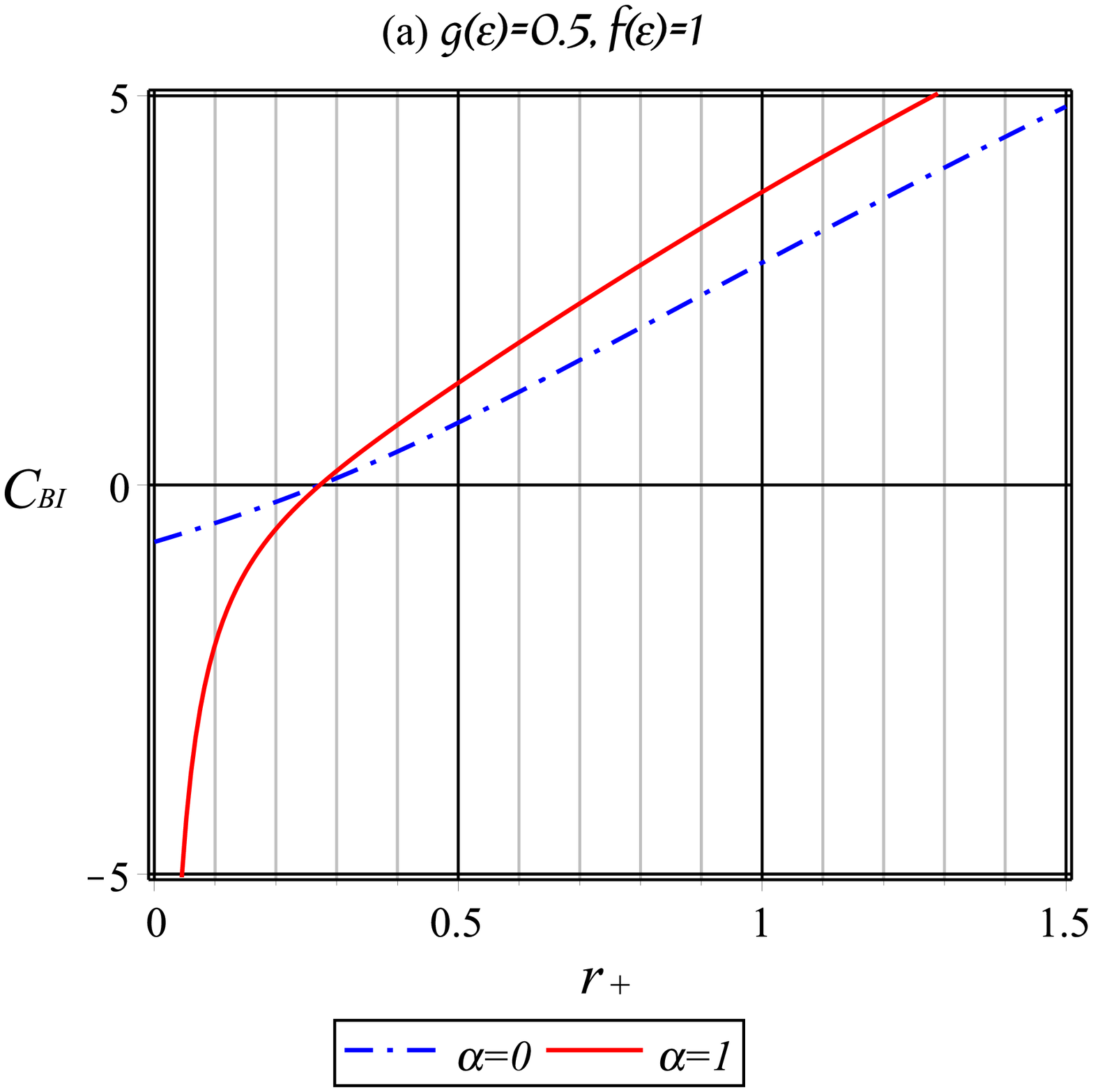}
\includegraphics[width=65 mm]{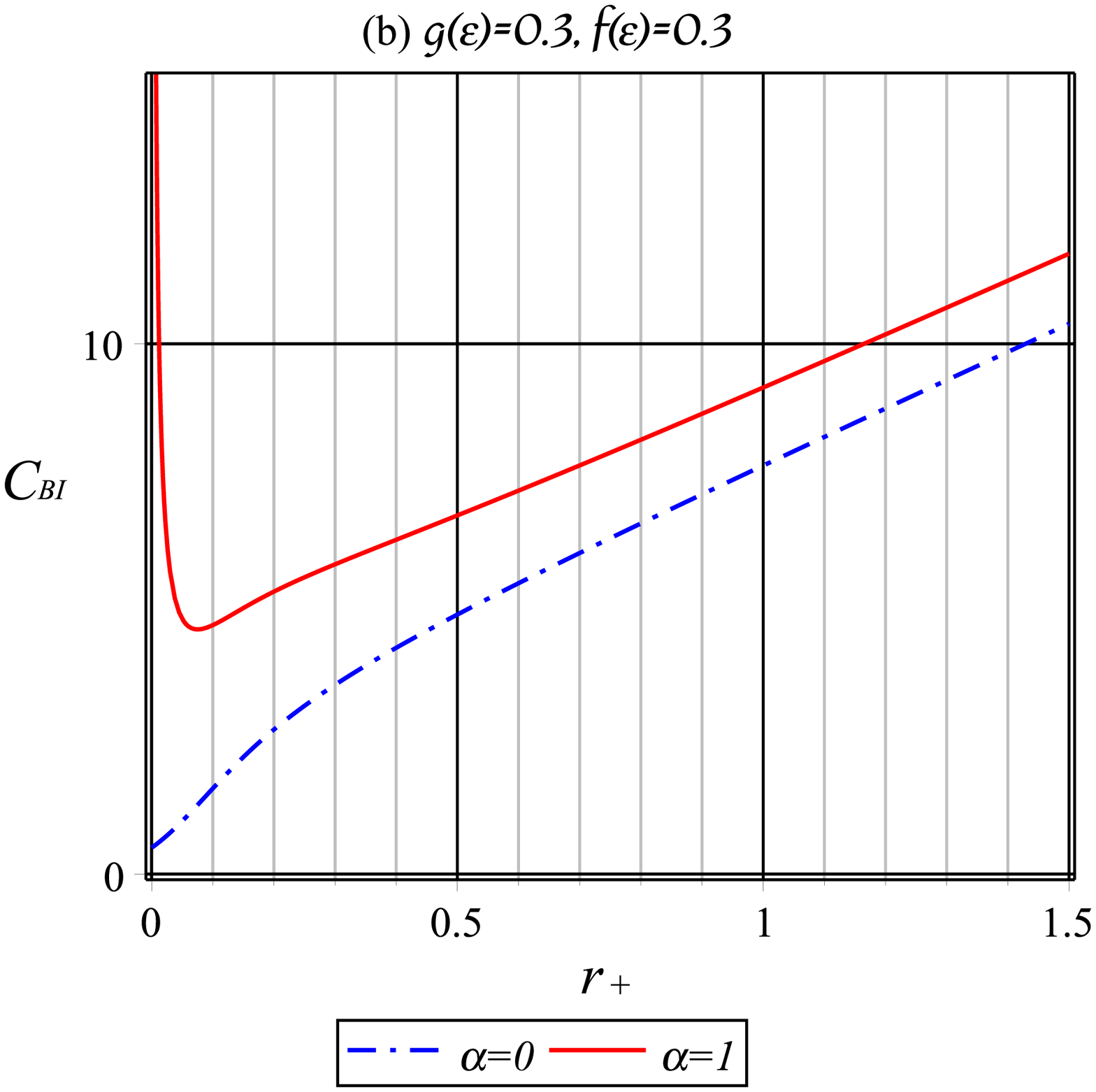}
 \end{array}$
 \end{center}
\caption{Corrected specific heat of BI model versus $r_{+}$ for $\Lambda=-1$, $c=c_{1}=l=1$, $b=1$, $q=1.2$, $m=2$, and $m_G=1$.}
 \label{fig4}
\end{figure}

In that case, the specific heat is corrected by the following term
\begin{equation}\label{CLnBI}
\delta C_{Q}=\frac{\varUpsilon_+\left[m^2_Gcc_1-4b^2r_+(\varUpsilon_+-1)-2\Lambda r_+ \right]} {2r_+\left[2b^2(\varUpsilon_+-1)-\Lambda \varUpsilon_+ \right]}
\end{equation}
Now, $C_{BI}=C_{Q}+\alpha\delta C_{Q}$ is the corrected specific heat. In plots of Fig. \ref{fig4} we can see the impact of logarithmic correction, and find that its effect is increasing the specific heat. We can see that, in the physical range, the black hole still is stable (Fig. \ref{fig4} (a)), but for the small values of rainbow parameters, (Fig. \ref{fig4} (b)) the asymptotic behavior happens near $r_{+}=0$, also we can see a minimum for the specific heat. We will discuss this issue in the next step.\\

\begin{figure}[h!]
 \begin{center}$
 \begin{array}{cccc}
\includegraphics[width=65 mm]{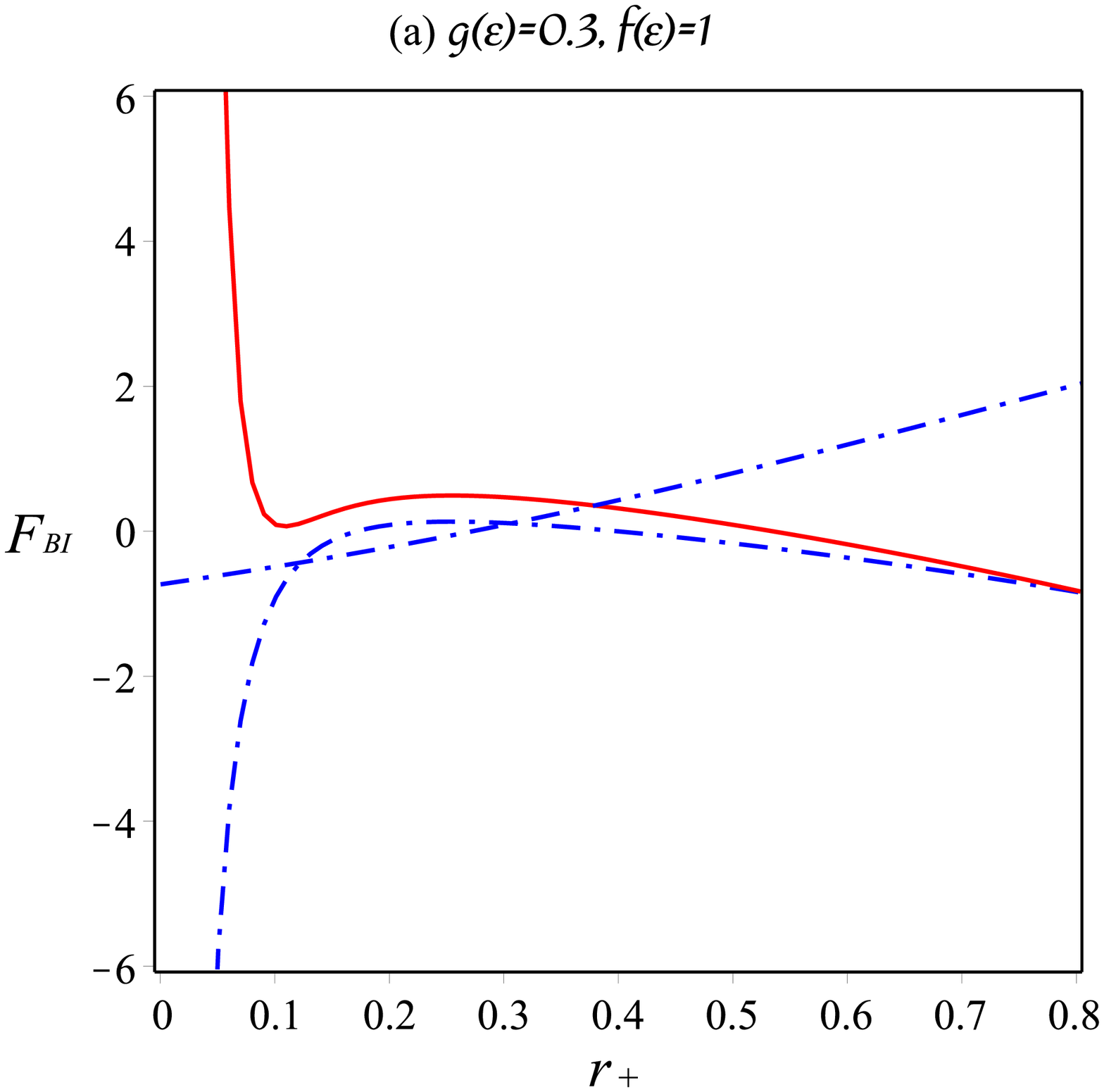}
\includegraphics[width=65 mm]{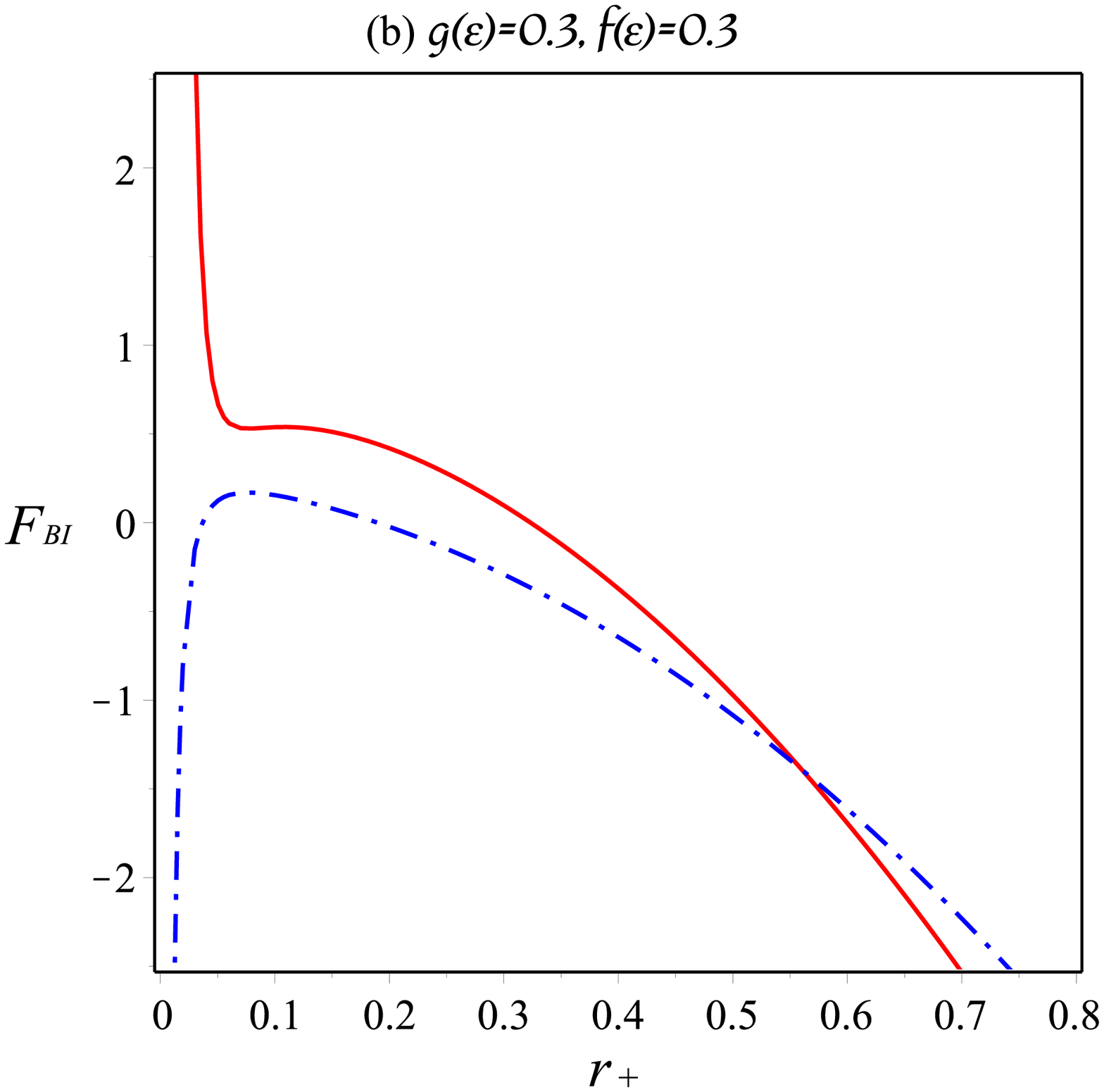}
 \end{array}$
 \end{center}
\caption{Corrected Helmholtz free energy of the BI model versus $r_{+}$ for $\Lambda=-1$, $c=c_{1}=l=1$, $b=1$, $q=1.2$, $m=2$, and $m_G=1$.}
 \label{fig5}
\end{figure}

Analyzing the Helmholtz free energy indicates that the minimum of specific heat is corresponding to the critical behavior of Helmholtz free energy where $dF/dr_{+}=d^{2}F/dr_{+}^{2}=0$. The Helmholtz free energy is given by,
\begin{equation}\label{FLnBI}
F_{BI}=-\int{SdT}=F_{0}+\alpha \delta F,
\end{equation}
in which
\begin{equation}\label{F0BI}
F_{0}=-\int{S_{0}dT}\approx-\frac{1}{4f(\varepsilon)g^{2}(\varepsilon)}\left(\frac{r_{+}^{2}}{2}+q_{\varepsilon}^{2}\ln{r_{+}}
+\frac{3q_{\varepsilon}^{2}}{b^{2}r_{+}^{2}}\right),
\end{equation}
while
\begin{equation}\label{F1BI}
\delta F\approx\frac{\bar{F}}{\pi f(\varepsilon)g(\varepsilon)},
\end{equation}
with
\begin{equation}\label{F11BI}
\bar{F}=\left(1-\ln{\frac{\pi r_{+}}{2g(\varepsilon)}}\right)\frac{r_{+}}{2}
+\frac{q_{\varepsilon}^{2}}{2r_{+}}\left(1+\ln{\frac{\pi r_{+}}{2g(\varepsilon)}}\right)
+\frac{q_{\varepsilon}^{4}}{8b^{2}r_{+}^{3}}\left(\ln{(\frac{2}{\pi r_{+}g(\varepsilon)})}-\frac{1}{3}\right).
\end{equation}
In Fig. \ref{fig5} we can see the behavior of Helmholtz free energy.  From Fig. \ref{fig5} (b) we can see a critical point corresponding to the minimum of specific heat.

\subsection{LO}
\begin{figure}[h!]
 \begin{center}$
 \begin{array}{cccc}
\includegraphics[width=65 mm]{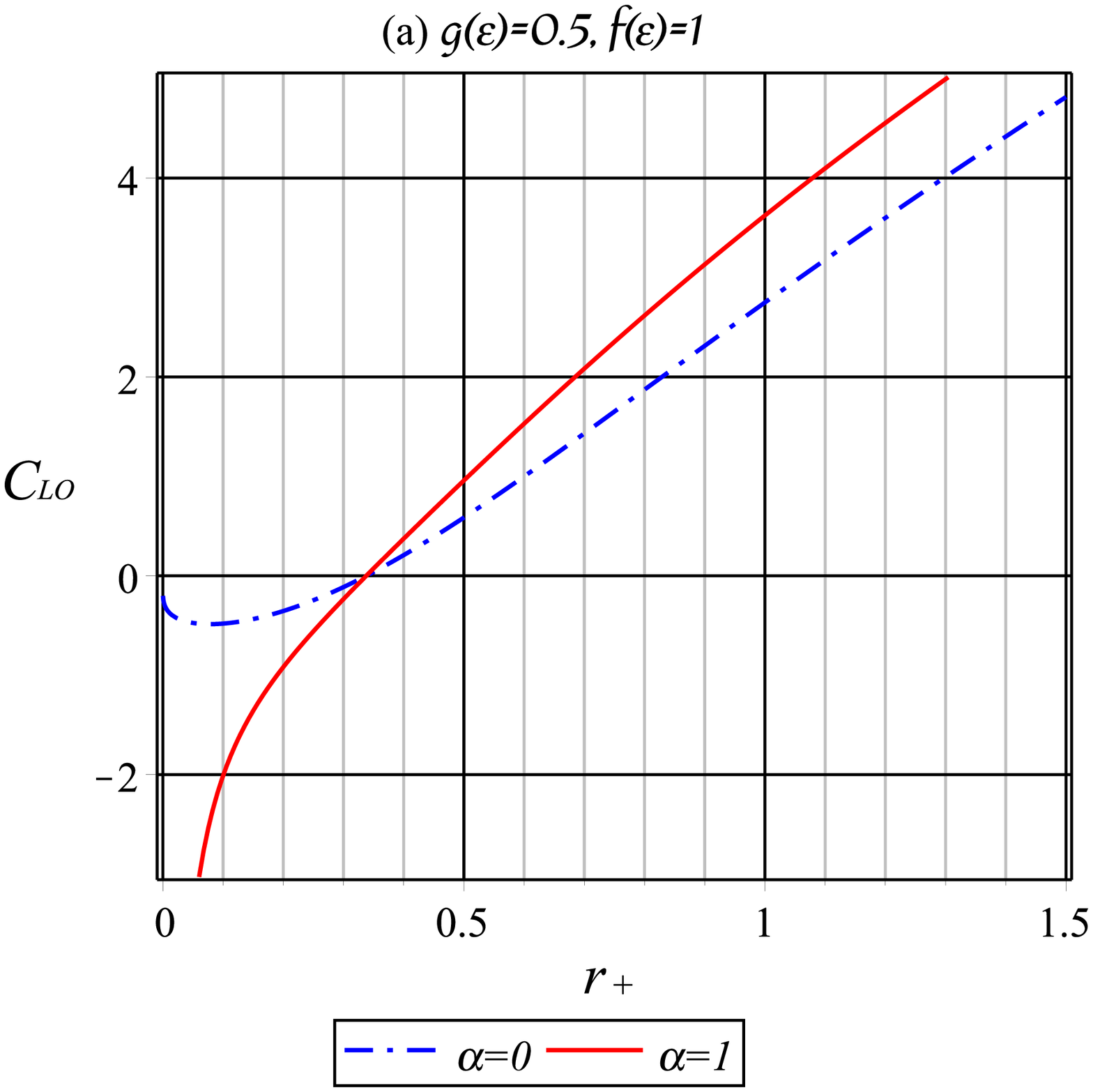}
\includegraphics[width=65 mm]{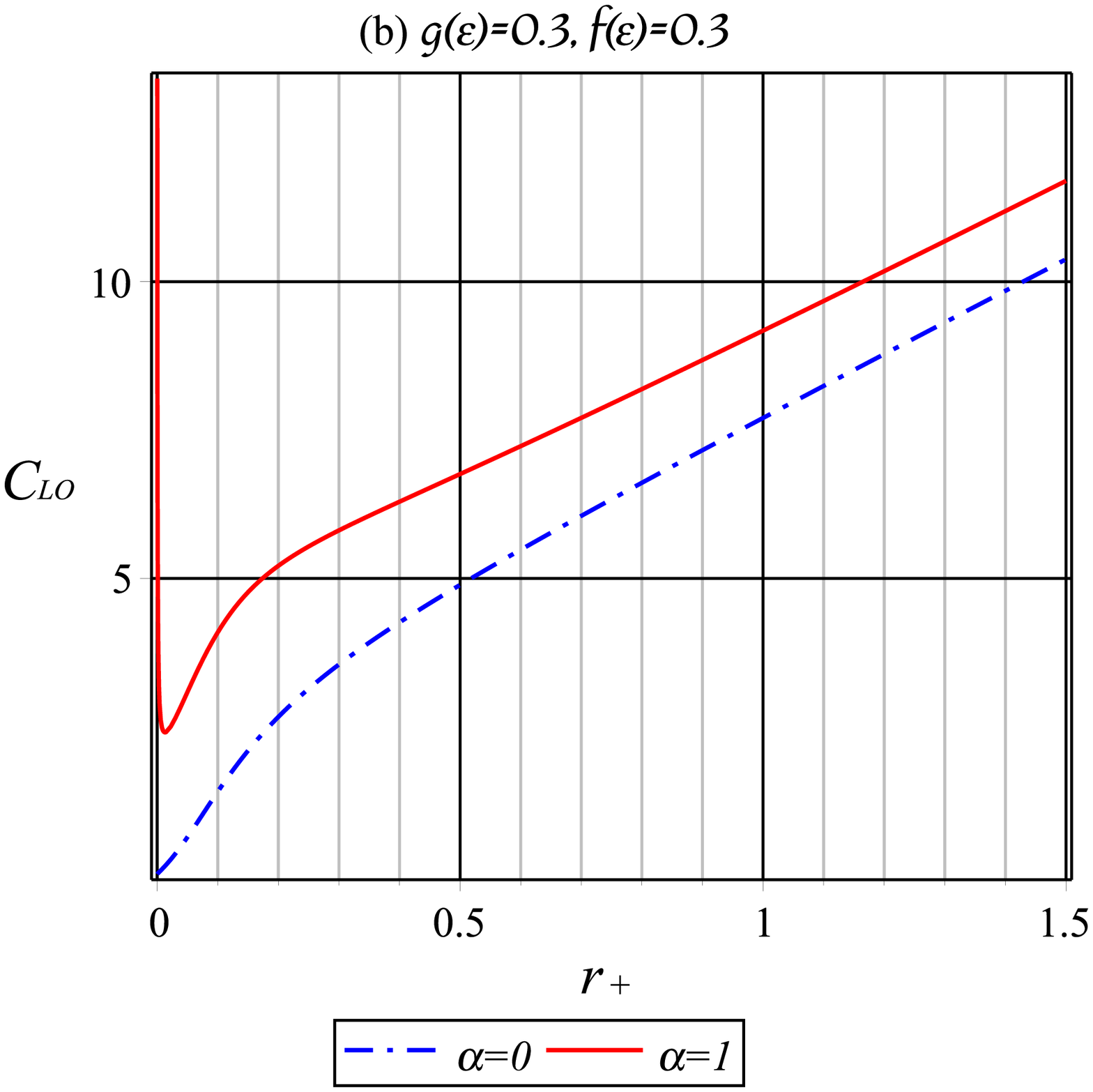}
 \end{array}$
 \end{center}
\caption{Corrected specific heat of the LO model versus $r_{+}$ for $\Lambda=-1$, $c=c_{1}=l=1$, $b=1$, $q=1.2$, $m=2$, and $m_G=1$.}
 \label{fig6}
\end{figure}

In the case of the LO model, the corrected specific heat is given by $C_{LO}=C_{Q}+\alpha\delta C_{Q}$, where
\begin{equation}\label{CLnLO}
\delta C_{Q}=\frac{ m^2_Gcc_1-2\Lambda r_+-8b^2r_+\left[ \varUpsilon_+-1-\ln\left(\frac{1+\varUpsilon_+}{2} \right) \right]} {2r_+\left[4b^2\ln\left(\frac{1+\varUpsilon_+}{2}\right)-\Lambda \right]}.
\end{equation}
Our numerical analysis represented by plots of Fig. \ref{fig6}, where we show the impact of logarithmic correction and find that its effect is increasing the specific heat, just like the previous model. We can see that in the physical range, the black hole still is stable (Fig. \ref{fig6} (a)), but for the small values of rainbow parameters (Fig. \ref{fig6} (b)) the asymptotic behavior occurs near $r_{+}=0$, which is the sign of the second-order phase transition. We can see similar behavior with the previous case, which also happens for the Helmholtz free energy.

\begin{figure}[h!]
 \begin{center}$
 \begin{array}{cccc}
\includegraphics[width=65 mm]{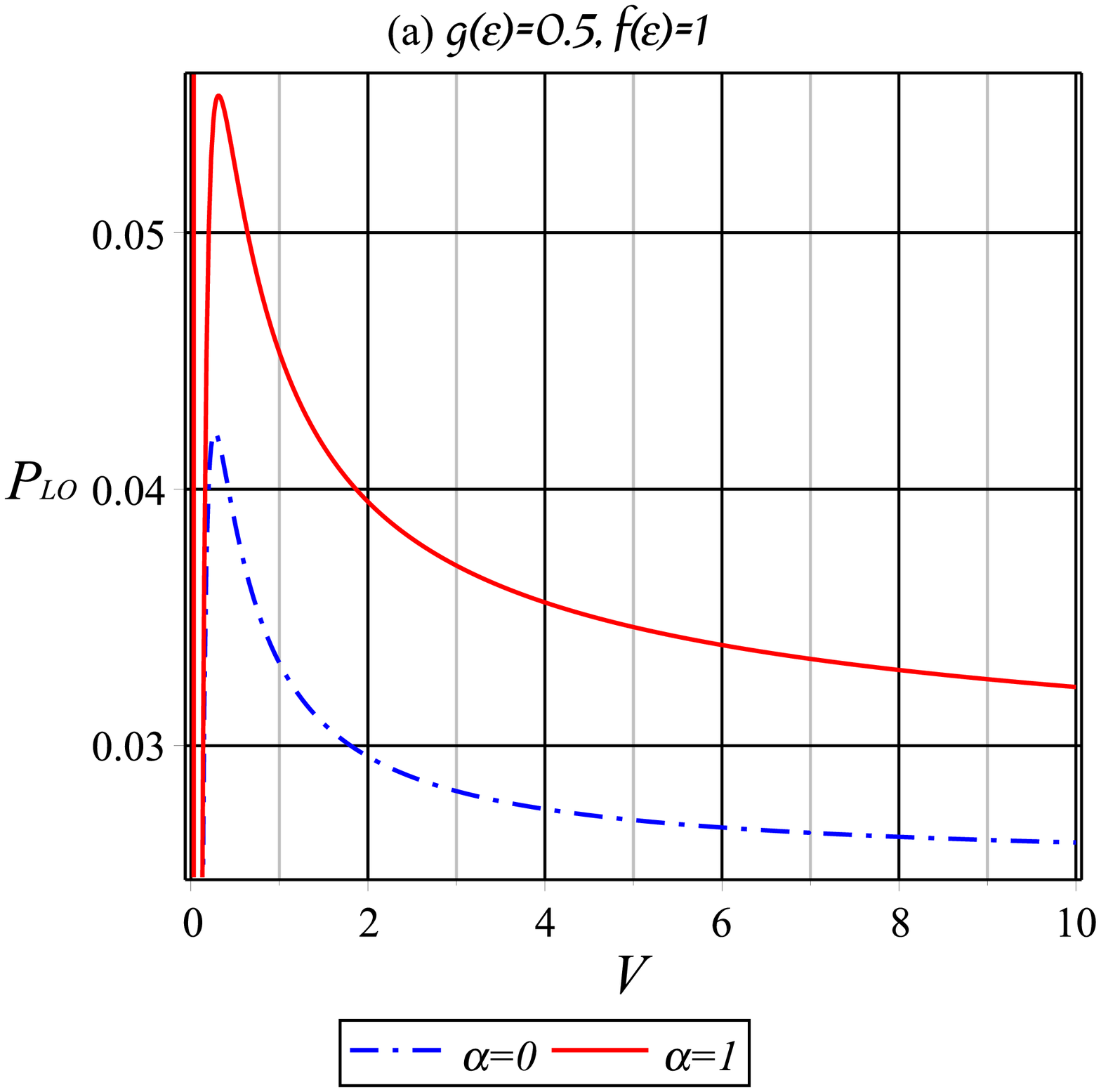}
\includegraphics[width=65 mm]{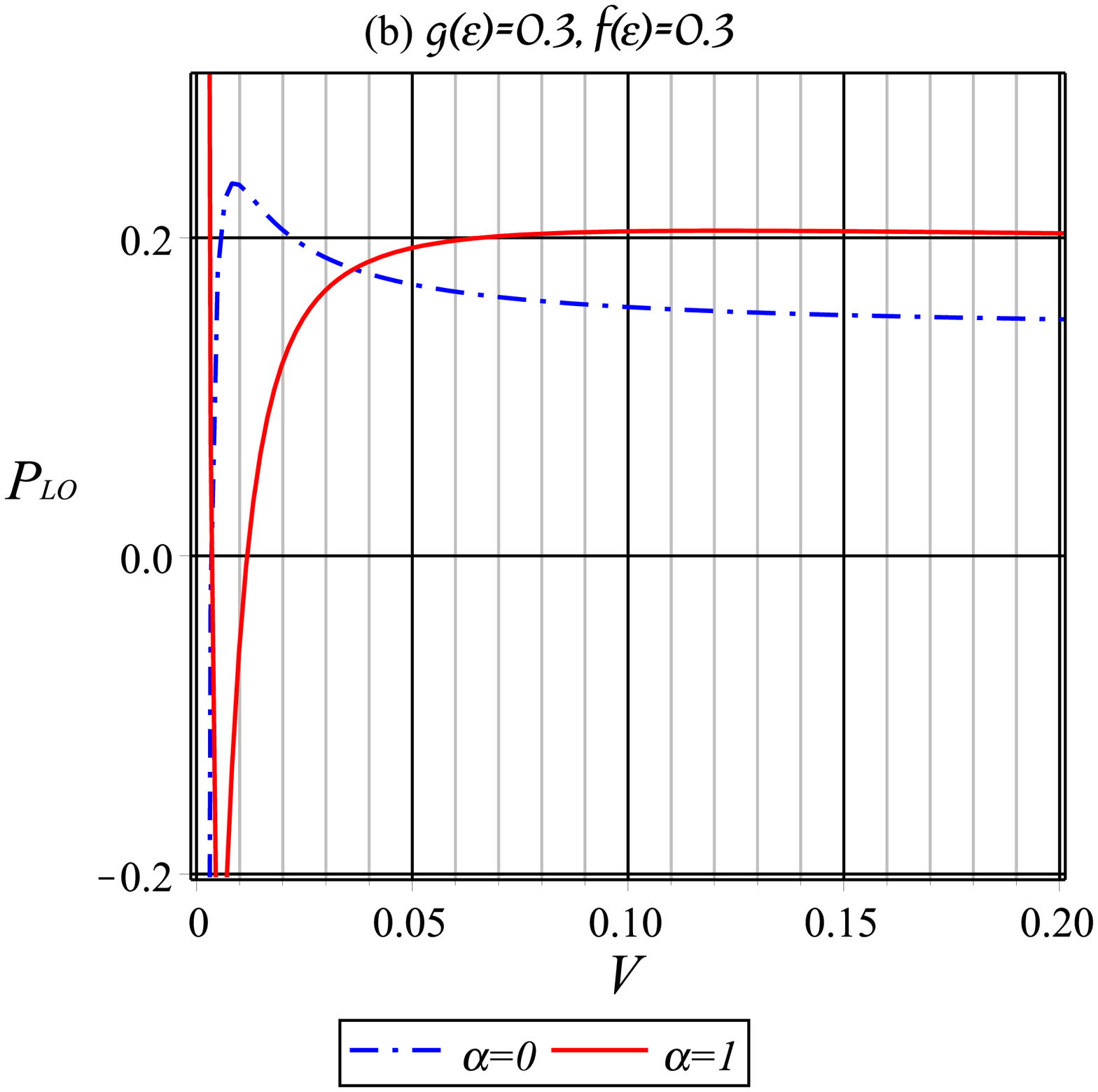}
 \end{array}$
 \end{center}
\caption{Pressure of the LO model versus $V$ for $\Lambda=-1$, $c=c_{1}=l=1$, $b=1$, $q=1.2$, $m=2$, and $m_G=1$.}
 \label{fig7}
\end{figure}

Having Helmholtz free energy, we can obtain pressure using
\begin{equation}\label{P}
P=-\left(\frac{\partial F}{\partial V}\right)_{Q}.
\end{equation}
We can use it to obtain $P-V$ diagram as illustrated by the plots of Fig. \ref{fig7}. We can see that consideration of logarithmic correction may increase or decrease the pressure depending on values of $f(\varepsilon)$ and $g(\varepsilon)$ as well as $V$. Fig. \ref{fig7} (a) shows that there is a maximum for the pressure at smallest volume, then it is a decreasing function of $V$. It means that at small volume (small radius) the pressure has a maximum value which happens at the phase transition point.

\subsection{EX}
\begin{figure}[h!]
 \begin{center}$
 \begin{array}{cccc}
\includegraphics[width=65 mm]{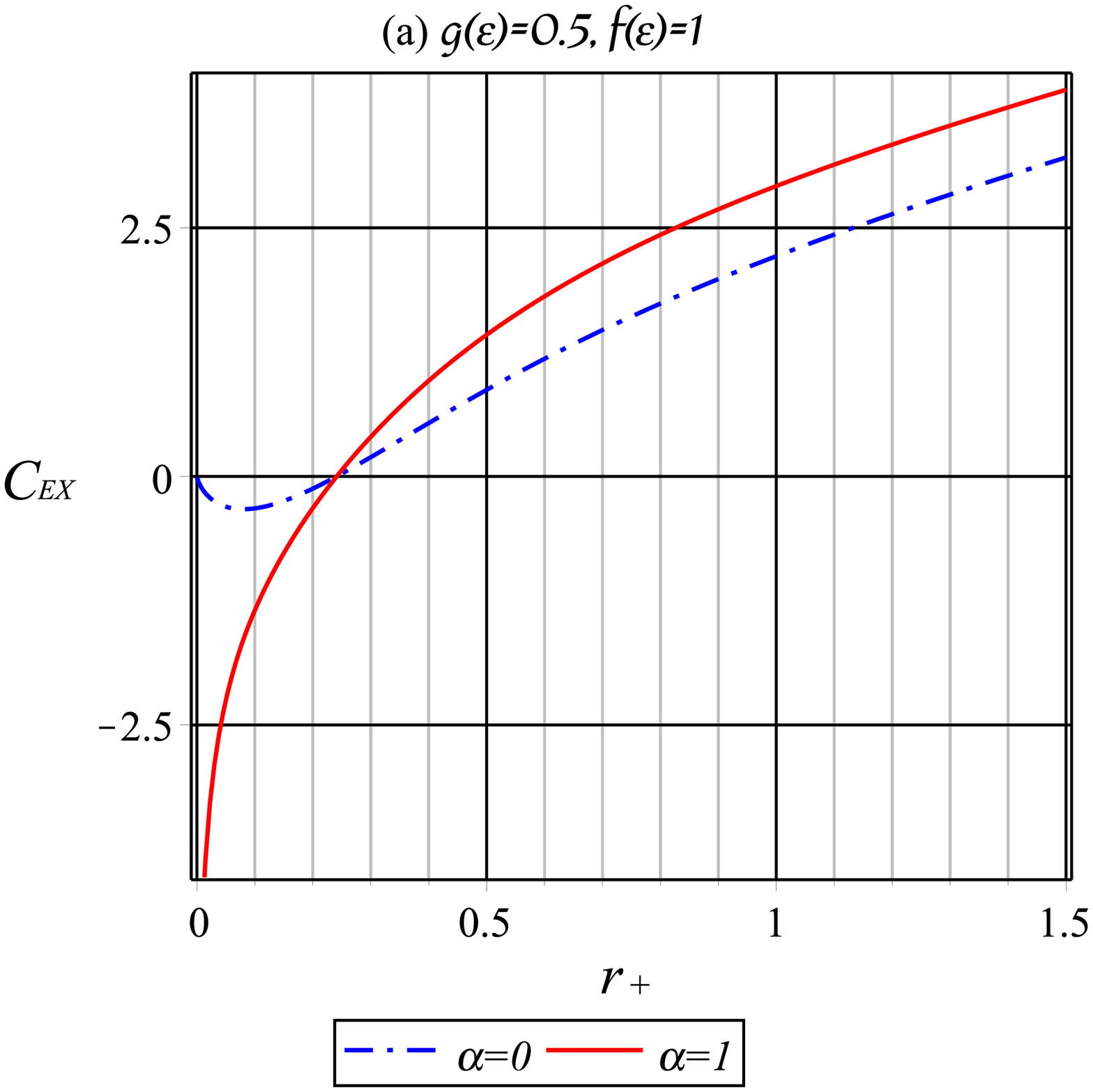}
\includegraphics[width=65 mm]{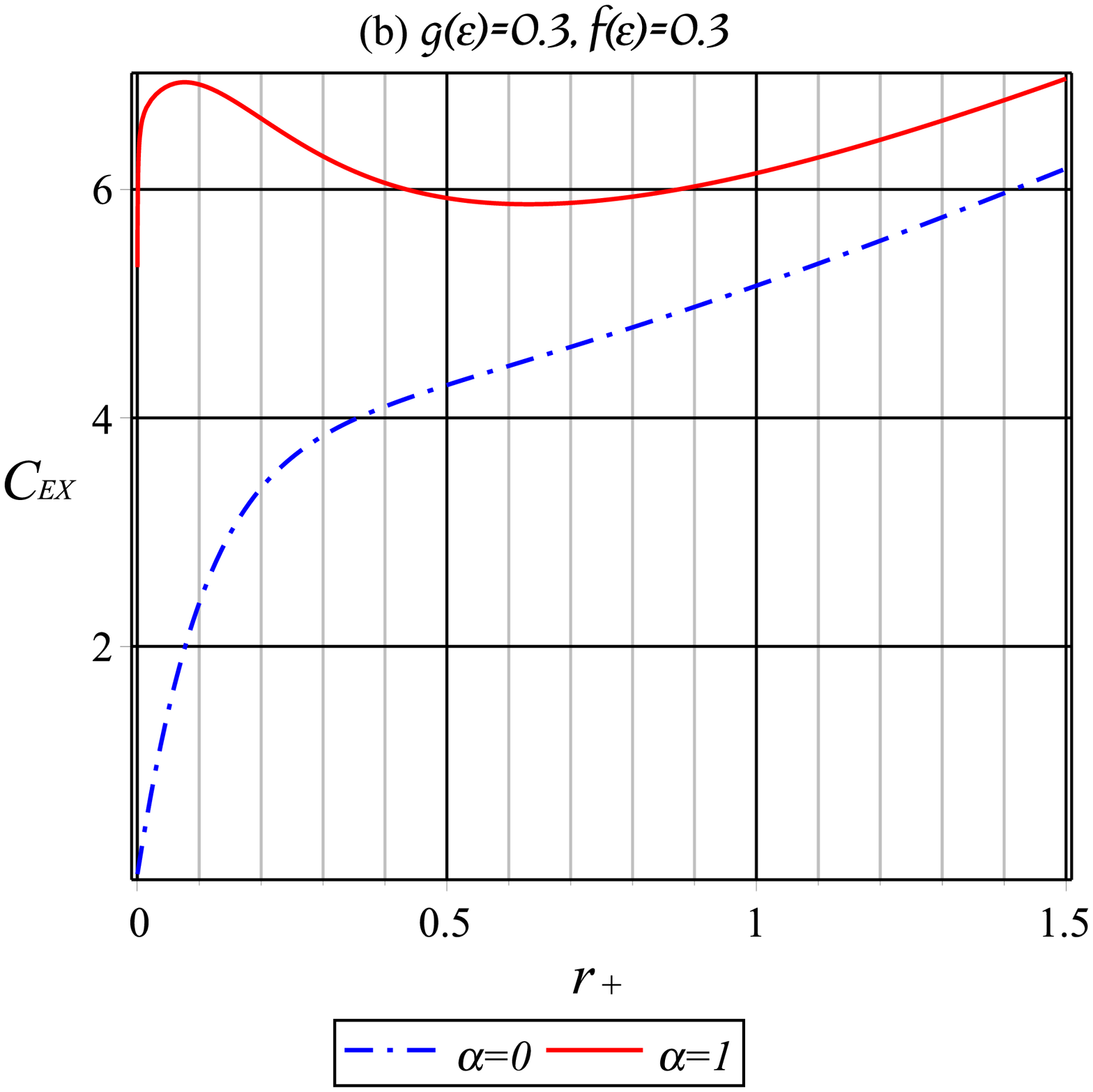}
 \end{array}$
 \end{center}
\caption{Corrected specific heat of the EX model versus $r_{+}$ for $\Lambda=-1$, $c=c_{1}=l=1$, $b=1$, $q=1.2$, $m=2$, and $m_G=1$.}
 \label{fig8}
\end{figure}

Now, we consider the final model, where the corrected specific heat is given by,
\begin{equation}\label{C-ln-EX}
C_{EX}=C_{Q}+\alpha\delta C_{Q},
\end{equation}
where
\begin{equation}\label{CLnEX}
\delta C_{Q}=\frac{m_G^2cc_1-(2\Lambda+b)r_+ +2bq_\varepsilon\left(\frac{1}{\sqrt{L_{W+}}}-\sqrt{L_{W+}} \right)} {2r_+b^2(e^{\frac{L_{W+}}{2}}-1)-4r_+\Lambda }.
\end{equation}
In Fig. \ref{fig8}, we draw the specific heat for different values of rainbow parameters. From Fig. \ref{fig8} (b) we can see something like Schottky anomaly in the presence of the logarithmic correction at small radii. Opposite to the previous cases, we can't see the second-order phase transition in this case.

%\newpage
\setcounter{equation}{0}
\section{Conclusion}
In this paper, we considered three-dimensional massive gravity theory, and solved the field equations in the presence of some nonlinear electrodynamics Lagrangian (Born-Infeld (BI), logarithmic (LO), and exponential (EX)). We obtained the exact black hole solutions of these theories in an energy-dependent circularly symmetric geometry. A notable point is that all the exact solutions reduce to their corresponding values in Einstein-Maxwell massive gravity theory when the nonlinearity parameter $b$ is chosen very large. Then, we discussed  thermodynamic properties of the novel nonlinearly charged energy-dependent black holes. We obtained the black hole temperature and discussed the zero-temperature limit to find extremal black hole solutions with horizon radius equal to $r_{ext}$. One can show that the physically reasonable black holes, having positive temperature, appear with $r_+>r_{ext}$ while those of negative temperature, named as unphysical black holes, occur for $r_+<r_{ext}$. We also obtained the black hole charge, entropy, and electric potential, as the other thermodynamic quantities. Then, by treating the black hole charge and entropy as the extensive thermodynamic quantities, we obtained the Smarr-type mass formula and used it to discuss the first law of black hole thermodynamics. Although, all of the thermodynamic quantities have been corrected in the presence of rainbow functions, the calculations confirm the validity of the thermodynamical first law for all the energy-dependent black holes, we just obtained. We analyzed the black hole specific heat and found that, in the absence of quantum corrections, there is only one point of first-order phase transition, which coincides with $r_{ext}$. The plots show that the black holes with horizon radii greater than $r_{ext}$ are locally stable. Also we found that, for tiny rainbow parameters, the black holes are entirely stable. Thermal fluctuations which may be interpreted as quantum effects modified the black hole entropy by a logarithmic term at leading order. It affect all thermodynamics quantities as discussed in this paper.  In the presence of quantum corrections, black hole entropy gets logarithmic correction term. It leads to modification of the first law of black hole thermodynamics. We have shown that, in the presence of quantum corrections, BI and LO black holes experience second-order phase transition too. Study of thermodynamic geometry and P-V criticality of the nonlinearly charged rainbow black holes obtained here, are suggested as the subjects of future works.\\\\
\textbf{Acknowledgement}: The first author thanks the Razi University Research Council for official supports of this work.\\\\

\end{document}